\documentstyle[aps,prb,preprint,tighten,epsf,eqsecnum]{revtex}

\newcommand{\calH}{{\cal H}}
\newcommand{\calA}{{\cal A}}
\newcommand{\calAdag}{{\cal A}^{\dagger}}
\newcommand{\calG}{{\cal G}}

\begin{document}

\title{Electron Spin Resonance in $S=1/2$ antiferromagnetic chains}

\author{Masaki Oshikawa$^1$ and Ian Affleck$^2$\footnote{
On leave from {\it
Department of Physics and Astronomy and Canadian
Institute for Advanced
Research, The University of British Columbia,
Vancouver, B.C., V6T 1Z1, Canada}}
}
\address{
$^1$Department of Physics, Tokyo Institute of Technology,
Oh-okayama, Meguro-ku, Tokyo 152-8551, Japan \\
$^2$Department of Physics, Boston University,
Boston, MA 02215}

\date{August 13, 2001}

\maketitle

\begin{abstract}
A systematic field-theory approach to
Electron Spin Resonance (ESR) in the $S=1/2$ quantum antiferromagnetic
chain at low temperature $T$
(compared to the exchange coupling $J$) is developed.
In particular, effects of a transverse staggered field $h$ and an
exchange anisotropy (including a dipolar interaction) $\delta$
on the ESR lineshape are discussed.
In the lowest order of perturbation theory, the linewidth is given
as $\propto Jh^2/T^2$ and $\propto (\delta/J)^2 T$, respectively.
In the case of a transverse staggered field, the perturbative
expansion diverges at lower temperature;
non-perturbative effects at very low temperature are discussed
using exact results on the sine-Gordon field theory.
We also compare our field-theory results with
the predictions of Kubo-Tomita theory
for the high-temperature regime, and discuss the crossover
between the two regimes.
It is argued that a naive application of the standard Kubo-Tomita
theory to the Dzyaloshinskii-Moriya interaction gives an incorrect
result.
A rigorous and exact identity on the polarization dependence
is derived for certain class of anisotropy,
and compared with the field-theory results.
\end{abstract}

\pacs{76.30.-v,11.10.Kk,75.10.Jm}

\section{Introduction}

Quantum spin chains have been studied extensively
for both their experimental and theoretical interests.
Among many experimental methods of investigation, Electron Spin
Resonance (ESR) is unique for its high sensitivity to
anisotropy.
While the theory of ESR has been
studied\cite{KuboTomita,MK:Ferro,MK:AF,ESRspindiff}
for a long time,
there remain important open problems, especially for
strongly interacting systems.
One of the problems is that, generally one has to
make a crucial assumption about the lineshape
at some point during the calculation.
As we will demonstrate, such an assumption could be
incorrect in some cases although it might have been
taken for granted in the literature.
In addition, in an actual calculation
one has to calculate various correlation functions.
Traditionally, crude approximations such as the high-temperature
approximation, the classical spin approximation and
the decoupling of the correlation functions are used.
However, these approximations break down when the
many-body correlation effects are strong.
As a consequence, rather little has been understood about ESR
when many-body correlations become important.
Even in the cases which were believed
to be understood with the existing theories,
there appear to be subtle problems.

In this paper, we study ESR in $S=1/2$
quantum spin chains in the ``one-dimensional critical region''
where the temperature $T$ is sufficiently small compared
to the characteristic energy of the exchange interaction $J$
(but $T$ is still large compared to three-dimensional ordering
temperature or spin-Peierls transition temperature.)
We stress that ESR in such a region is essentially a
many-body problem.
Here, many of the traditional theoretical techniques
lose their validity.
Instead, $(1+1)$-dimensional field theory should describe the universal,
low-energy/large-distance behavior.
Our main purpose in the present paper is to develop a new
approach to ESR based on field theory (bosonization) methods.
At least for several simple cases (which are of experimental interest)
we are able to formulate the problem in terms of the
systematic Feynman-Dyson perturbation theory, avoiding
previously made ad hoc assumptions.
We will study several consequences of our theory for
two types of perturbations of
the one-dimensional $S=1/2$ Heisenberg antiferromagnet:
a staggered field, and an exchange anisotropy (or dipolar interaction).
When the effect of the anisotropy is small, the ESR lineshape
is shown to be Lorentzian up to a possible small smooth background;
the width and the shift of the Lorentzian peak are given
perturbatively.
In one dimension, it was argued that the diffusive spin dynamics
leads to a non-Lorentzian lineshape, which is indeed observed in
the $S=5/2$ antiferromagnetic chain TMMC\cite{ESRspindiff}.
However, our results imply that the argument
does not apply to the present
case of the $S=1/2$ antiferromagnetic chain at low temperature.

In a compound with a low crystal symmetry permitting a staggered component
of the gyromagnetic tensor or a staggered
Dzyaloshinskii-Moriya (DM) interaction,
an effective staggered field is also produced by the applied uniform field.
The staggered field corresponds to a relevant operator in the
Renormalization Group sense, and is related to the field-induced
gap phenomenon recently found in several quasi-one
dimensional $S=1/2$ antiferromagnets\cite{Dender1,Cubenz-PRL,Yb4As3,Feyerherm}.
Since it is a relevant operator, one may expect that its effect
is enhanced at lower temperatures.
Indeed, we find that the staggered field contributes to the linewidth
proportionally to $h^2/T^2$ where $h$ is the magnitude of the
staggered field.
We propose this as an explanation of
the peculiar low-temperature behavior\cite{Okuda}
found in ESR on Cu Benzoate nearly 30 years ago.
Moreover,
we propose that the sharp resonance found
at very low temperature\cite{Oshima:AFMR},
which was understood as a signature of a three-dimensional N\'{e}el ordering,
may well be understood in a purely one-dimensional framework
based on sine-Gordon field theory.

On the other hand, dipolar interactions or exchange anisotropies
are present in virtually any real material.
We find that their contribution to the linewidth is proportional to $T$,
which appears to be consistent with existing experimental data
on several quasi one dimensional $S=1/2$ antiferromagnet
such as CuGeO$_3$, KCuF$_3$ and NaV$_2$O$_5$.

Basic ideas and some of the results in the present paper
were presented briefly in Ref.~\onlinecite{ESR-PRL}.
This paper is organized as follows.
In Section~\ref{sec:esr},
we briefly review the basics of ESR
in interacting spin systems, including
a few (apparently) new results,
namely an exact and rigorous identity
on the polarization dependence, and the relation between
the Kubo-Tomita\cite{KuboTomita} and the
Mori-Kawasaki\cite{MK:Ferro,MK:AF}
theories.
In Sections~\ref{sec:ft} and ~\ref{sec:self-energy} we develop a
new framework for studying ESR in quantum spin chains, based on
field theory methods and, in particular, the Dyson formula
expressing the Green's function for a scalar field in terms of the
self-energy. It is applied in Sections~\ref{sec:anis-z},
\ref{sec:anis-x}, \ref{sec:xxz} and ~\ref{sec:stag} to
systems with an exchange anisotropy (or dipolar interaction)
or a transverse staggered field. (The case of an exchange
anisotropy with the axis parallel to
the field turns out to be easier to treat and not to require
the self-energy formalism.  Therefore it is treated first, in
Section~\ref{sec:anis-z}.) In Section~\ref{sec:hight}, we
compare our results to those in the high-temperature regime
obtained with the previous approach. Section~\ref{sec:conc} is
devoted to conclusions. Appendix A contains an alternative
derivation of an old formula for the width/shift first derived by
Mori and Kawasaki \cite{MK:Ferro,MK:AF}.

\section{Electron Spin Resonance}
\label{sec:esr}

\subsection{Definition of the problem}
\label{sec:defesr}

A single spin in a magnetic field $H$ has energy levels separated by
the Zeeman energy $E_Z = g \mu_B H$.
If an electromagnetic wave of angular frequency $\omega$
is applied to such a system, resonant absorption
occurs when $\hbar \omega = E_Z$ and the polarization
(direction of the oscillating magnetic field) is perpendicular to
the static field.
When the spins are coupled by interactions,
the physics is of course not that simple.
However, generally some resonant absorption occurs also in
the interacting system.
This is the phenomenon of ESR which we study in the present paper.
In an interacting system, it is also possible to observe
absorption of the electromagnetic wave polarized parallel to
the static magnetic field (so called Voigt configuration.)
In this paper, we focus on the standard (Faraday) configuration,
which measures the absorption of the electromagnetic wave
polarized perpendicular to the static magnetic field.

Assuming that the absorption can be described by linear
response theory, the absorption intensity $I(\omega)$ per volume
for the radiation linearly polarized in the $\alpha \perp z$ direction
is given by
\begin{equation}
I(\omega) = \frac{{H_R}^2 \omega}{2} \chi''_{\alpha \alpha}(q=0,\omega),
\label{eq:defI}
\end{equation}
where $H_R$ is the amplitude of the radiation and
$\chi''$ is the imaginary part of the dynamical
magnetic susceptibility.
$\chi''$ is related to the retarded Green's function $\calG^R$ as
\begin{equation}
\chi''_{\alpha \beta}(q,\omega) =
    - {\rm Im}\calG^R_{\alpha \beta} (q, \omega),
\end{equation}
where $\calG^R_{\alpha \beta}$ is defined by
\begin{equation}
\calG^R_{\alpha \beta} (q, \omega) = - i
\int_0^{\infty} dt \sum_x
    \langle [ S^{\alpha}(x,t) , S^{\beta}(0,0) ] \rangle
    e^{- i q x + i \omega t}
\label{eq:defgr}
\end{equation}
where $\langle \ldots \rangle$ is the statistical average
at temperature $T$.
In most experiments, the applied electromagnetic wave is
typically in the microwave regime, and its wavelength is
very large compared to all relevant length scales in
the antiferromagnet since the spin-wave velocity is much less
than the speed of light.
Thus, in ESR, the dynamical susceptibility is measured at essentially
zero momentum $q=0$.
ESR probes the dynamics of the system only
at the special momentum $q=0$, in contrast
to neutron scattering
which can be used to scan momentum space.
However, as we will explain below,
there is an interesting feature at the special momentum $q=0$.
Together with the relatively easy availability of
highly precise data, ESR offers a unique insight into
magnetic systems which would be difficult to obtain with
other experimental methods.

A remarkable feature of ESR is that,
if the Hamiltonian of the system (apart from the Zeeman term)
is isotropic (ie. $SU(2)$ symmetric),
the resonance is still at the Zeeman energy and completely sharp,
as if there is no interaction at all.
This result can be deduced rather easily from the equation
of motion, as we will show in the following.
Throughout this paper, we take the direction of the static applied
field as the $z$-axis.
Let us consider the total Hamiltonian
\begin{equation}
\calH = \calH_0 + \calH_Z,
\end{equation}
where $\calH_Z = - H \sum_j S^z_j$ is the Zeeman term
and $\calH_0$ is the exchange Hamiltonian which is assumed to be
$SU(2)$ symmetric.
We choose units so that $\hbar = g \mu_B = 1$
except where explicitly mentioned otherwise;
these constants can be recovered by dimensional analysis.
We also assume $H$ is much smaller than the exchange coupling $J$
(ie. energy scale of $\calH_0$).
As we have mentioned above, in ESR the electromagnetic wave is
coupled to the $q=0$ component of the spin operators, namely
the total spin operators $S^{\alpha} = \sum_j S^{\alpha}_j$.
The Heisenberg equation of motion for $S^+ = S^x + i S^y$ reads
\begin{equation}
\frac{d S^+}{dt} = i [ \calH , S^+] =  i [ \calH_Z , S^+] = - i H S^+ ,
\label{eq:eqofM}
\end{equation}
because $\calH_0$ commutes with $S^+$ due to the $SU(2)$
symmetry of $\calH_0$.
It follows that $S^+(t) = S^+ e^{-iHt}$, and consequently
$\chi^{+-}(0,\omega) \propto \delta(\omega - H)$.
This means that the resonance is completely sharp,
and located exactly at the Zeeman energy.
Namely, this resonance has the lineshape
identical to ESR in a single (non-interacting) spin in spite of an arbitrary
strong exchange interaction.
On the other hand, the
absorption intensity is generally affected
by the exchange interaction $\calH_0$.
For example, in a spin-gap system at zero temperature, the absorption intensity
is zero if the applied field $H$ is smaller than the gap.

As we have seen, the completely sharp resonance is related to the
$SU(2)$ symmetry of the exchange Hamiltonian $\calH_0$. While it
is natural that symmetries of the system are important in
determining the dynamics of the system, the present situation is
rather unique, for the $SU(2)$ symmetry is explicitly broken down
to $U(1)$ by the applied static field but is still essential in
ESR. This peculiar feature is related to the fact that the applied
field couples to the total magnetization $S^z = \sum_j S^z_j$,
which is a generator of the global $SU(2)$ symmetry and is
conserved under $\calH_0$.
Since the total magnetization and
Hamiltonian are simultaneously diagonalizable, 
the applied field does not change the eigenstates of
the system, if they are classified by $S^z$. The only effect of the static
applied field is to shift the energy levels of the eigenstates;
the shifted energy levels still reflects the $SU(2)$ multiplet
structure. This kind of ``weak'' symmetry breaking by one of the
symmetry generators preserves some structures of the fully
symmetric system. In ESR of an isotropic system $\calH_0$, the
$SU(2)$ symmetry is only weakly broken and is essential in
determining the ESR spectrum.

A similar application of the concept of weakly broken global symmetry
was also exploited recently by S.~C.~Zhang\cite{SO5} in his
$SO(5)$ theory of high-$T_c$ superconductivity.
Namely, in the $SO(5)$ theory,
the most important terms in the effective Hamiltonian are $SO(5)$ symmetric one
and the chemical potential couples to one of the generators
of the global $SO(5)$ symmetry.
The so-called $\pi$-excitation in this context
is a sharp resonance which is similar to ESR in isotropic
spin systems.

In real magnetic systems, there are
various types of anisotropy, such as
the dipolar interaction.
Let us write the total Hamiltonian as
\begin{equation}
 \calH = \calH_0 + \calH' + \calH_Z,
\label{eq:totalH}
\end{equation}
where $\calH'$ is the symmetry-breaking perturbation.
Throughout this paper, we assume the interaction to be nearly isotropic,
namely that $\calH'$  is small compared to the other terms
$\calH_0$ and $\calH_Z$.
Once the perturbation $\calH'$ is added, the argument
leading to the delta-function resonance at the Zeeman energy
breaks down.
Thus, in general, the addition of $\calH'$ causes changes
in the lineshape, such as
a broadening and a shift of the resonance.
The main theoretical problem is then to calculate the absorption
spectrum for the given Hamiltonian $\calH$ and other conditions
such as the temperature of the system.

\subsection{Previous theories}

The existing approaches to ESR, such as those of
Kubo and Tomita~\cite{KuboTomita} and
of Mori and Kawasaki~\cite{MK:Ferro,MK:AF}
were developed mainly during 1950s-60s.
Here we summarize briefly,
a part of those achievements which is closely
related to our analysis.

When the isotropic exchange interactions between spins are weak, namely
$\calH_0$ is much smaller than $\calH_Z$, the lineshape is generally
expected to be Gaussian.
On the other hand, once the anisotropy $\calH'$ is present,
strong isotropic exchange interactions $\calH_0$
between spins affects the ESR lineshape,
even though it does not break the $SU(2)$ symmetry
by itself.
In the presence of the strong interaction ($\calH_0 \gg \calH_Z$),
which applies to the problem considered in this paper,
the lineshape is generally expected to be Lorentzian.
(On the other hand, the lineshape is believed to be neither Gaussian
nor Lorentzian, when the spin diffusion is dominant.~\cite{ESRspindiff})
The effect of the isotropic exchange interactions on the lineshape
has been traditionally called Exchange Narrowing.
We emphasize that ESR in such an interacting
spin system probes the collective motion
of the many-body system.
In this paper, we focus on this limit of strong isotropic exchange
interaction, while other cases have been discussed
previously as well \cite{KuboTomita,MK:Ferro,MK:AF}.

For the case of the Lorentzian lineshape,
Mori and Kawasaki\cite{MK:Ferro} proposed a formula, which we call
the MK formula, for the linewidth $\eta$:
\begin{equation}
\eta = \frac{1}{2 \chi_u H} {\rm Im}{[- G^R_{\calA \calAdag}(\omega=H)]},
\label{eq:MKwidth}
\end{equation}
where $\chi_u$ is the magnetic susceptibility and
$G^R_{\calA \calAdag}(\omega)$ is the Fourier transform of the
{\em unperturbed} retarded
Green's function
\begin{equation}
G^R_{\calA \calA^{\dagger}}(t) =
 -i \theta(t) \langle
 [ \calA (t) , \calA^{\dagger}(0) ] \rangle_0 ,
\label{eq:defGRA}
\end{equation}
where $\langle \ldots \rangle_0$ is the expectation value under
the unperturbed Hamiltonian $\calH_0 + \calH_Z$,
$\theta(t)$ is the step function, and $\calA$ is defined by
the commutator
\begin{equation}
    \calA = [ \calH' , S^+].
\label{eq:defA}
\end{equation}
In this paper, $\calG$ refers to a full Green's function
calculated using
the Hamiltonian including the perturbation $\calH'$,
while $G$ denotes the unperturbed Green's function evaluated
in the absence of the perturbation.
Both kinds of Green's functions ($\calG$ and $G$)
should be evaluated including the Zeeman term $\calH_Z$,
in the original spin chain context.
However, as we will see in Section~\ref{sec:ft},
in the effective field theory,
the Zeeman term is absorbed by a momentum shift.
Thus, the Green's functions in the effective field theory
will be defined without explicitly including the Zeeman term.

In addition to the broadening,
the perturbation $\calH'$ also causes a shift of the resonance energy;
the shift is given by
\begin{equation}
\Delta \omega  = \frac{-1}{2 \chi_u H} \left\{
     \langle [\calA, S^-] \rangle
- {\rm Re} G^R_{\calA\calA^{\dagger}}(\omega=H)
                      \right\}.
\label{eq:MKshift}
\end{equation}
This formula for the shift is slightly different from the one
given in the original paper\cite{MK:Ferro}.
We believe that ours is the correct one in the lowest order of 
perturbation theory.

The derivation of the MK formulae in the original paper
seems somewhat involved, and it is not clear to us what
assumptions are necessary to prove them.
However, we found that the MK formulae is indeed
exact in the lowest order of the perturbation theory,
{\em if the (single) Lorentzian lineshape is assumed.}
Explicitly speaking, we must assume:
\begin{equation}
\calG^R_{S^+ S^-}(\omega) = \frac{2 \langle S^z \rangle}{\omega - H - \Sigma},
\label{eq:lor}
\end{equation}
where $\Sigma$ is a smooth function of $\omega$ near $\omega =H$.
Regarding $\Sigma$ as a constant near the resonance,
we obtain a Lorentzian line-shape.  [Setting $\Sigma =0$ in
Eq. (\ref{eq:lor}) gives the exact result for the isotropic case
${\cal H}'=0$.]
The simple, and possibly new, alternative derivation using the equation of
motion is presented in Appendix~\ref{sec:deriveMK}.

On the other hand, Kubo and Tomita\cite{KuboTomita} studied ESR using
a somewhat different formulation.
For the case of Lorentzian lineshape, their theory gives the
following formula for the linewidth, at high temperature:
\begin{equation}
\eta \sim \frac{1}{| J|}
\frac{ \langle \calA \calAdag \rangle}{
        \langle S^+ S^- \rangle },
\label{eq:KTwidth}
\end{equation}
where the expectation value is the {\it static} correlation function.
We shall call this the KT formula in this paper.
We could not find in the literature how the two formulae
(\ref{eq:MKwidth}) and (\ref{eq:KTwidth}) are related.
On the other hand, if the KT formula~(\ref{eq:KTwidth})
for the Lorentzian lineshape is indeed valid
at high temperature, it must be consistent with the MK formula.
In fact, we have verified that
the KT formula~(\ref{eq:KTwidth}) follows from the high-temperature
limit of the MK formula~(\ref{eq:MKwidth}) with a certain assumption.
The derivation is given as follows.
Taking the Fourier transform of eq.~(\ref{eq:defGRA}), at temperature $T$,
\begin{equation}
G^R_{\calA \calAdag}(\omega) =
- \frac{i}{Z}
\int_0^{\infty} dt e^{i\omega t}
{\rm Tr}{\left( [ \calA(t), \calAdag(0) ] e^{- (\calH_0 + \calH_Z)/T} \right)},
\end{equation}
where $Z = {\rm Tr}{e^{-(\calH_0+\calH_Z)/T}}$.
Expanding this up to the first order in $1/T$, we find,
\begin{eqnarray}
G^R_{\cal A \calAdag}(\omega) &\sim&
\frac{i}{T Z_{\infty}} \int_0^{\infty} dt
{\rm Tr}{\left( [ \calH_0 + \calH_Z , \calA(t)] \calAdag(0) \right)}
e^{i\omega t}
\nonumber \\
&=&
\frac{1}{T Z_{\infty}} \int_0^{\infty} dt
{\rm Tr}{\left( \frac{d \calA}{dt}(t) \calAdag(0) \right)} e^{i\omega t}
\nonumber \\
&=&
- \frac{1}{T Z_{\infty}} {\rm Tr}\left[{\calA(0)\calAdag(0)}\right]
- i \frac{\omega}{T Z_{\infty}} \int_0^\infty dt
{\rm Tr}\left[ {\calA(t)\calAdag(0)} \right] e^{i \omega t},
\end{eqnarray}
where the time evolution is defined with respect to the unperturbed
Hamiltonian $\calH_0 + \calH_Z$ and
$Z_{\infty} = {\rm Tr} 1$ is the partition function in
the infinite temperature limit.
The first term is real and does not contribute to the imaginary part.
If we assume that the dynamical correlation function at infinite
temperature ${\rm Tr}\left[{\calA^{\dagger}(t)\calA(0)}\right]$ decays
exponentially with the characteristic time $\tau_c \sim 1/J$,
the second term gives $- i \frac{\omega}{J T}
\langle {\calA^{\dagger}(0) \calA(0)} \rangle_{\infty}$,
where $\langle \rangle_{\infty}$ is the expectation value
at the infinite temperature and we use $\omega \ll J$.
We note that a similar assumption was made also in the
original derivation of eq.~(\ref{eq:KTwidth})
in the Kubo-Tomita paper.
Thus the MK formula~(\ref{eq:MKwidth}) reduces, in the
high-temperature limit, to
\begin{equation}
\eta \sim \frac{\langle \calA \calAdag \rangle_{\infty}}{2 \chi_u
T J} . \label{eq:KT2} \end{equation} Because $\langle S^+ S^-
\rangle \sim 2 \chi_u T$ in the high-temperature limit, this is
equivalent to the KT formula~(\ref{eq:KTwidth}). We note that,
because $\tau_c \sim 1/J$ is valid only as an order-of-magnitude
estimate at best, the KT formula has the uncertainty of an overall
constant factor.

Recently, a numerical approach to ESR in quantum spin chains
is also being developed\cite{Miya-ESR} by a direct calculation
of the dynamical susceptibility $\chi''(\omega)$.
Since it is based on an exact diagonalization of the full spectrum
of short chains,
it is restricted to rather short chain of up to 10 spins even for
$S=1/2$, making finite size effects rather severe.
On the other hand, the direct numerical calculation is applicable
at any temperature.
In contrast, the field theory approach, which we will develop
in the present paper, is valid only at low temperatures while
it is based on the thermodynamic limit.
Thus they are complementary to each other.

We remark that some results quite closely related to ours were 
derived by Giamarchi and Millis\cite{Giamarchi,Giamarchi2}
 in their work on the ac  
conductivity of a Tomonaga-Luttinger (TL)  liquid.  We will comment on the 
connections with our work later.
\subsection{Polarization dependence}
\label{sec:polari}

When observing ESR in the Faraday configuration,
the polarization of the electromagnetic wave is perpendicular
to the direction of the static magnetic field, which we take
as the $z$-axis.
There are still two independent possible polarizations;
the linear polarization can take any direction in the $xy$-plane.
Except when the total Hamiltonian $\calH$ is invariant
under a rotation about the $z$ axis, the absorption spectrum
generally depends on the polarization.
Within the linear response theory, the dependence comes
from the difference between the dynamical susceptibility
$\chi''_{xx}(0,\omega) \neq \chi''_{yy} (0, \omega)$.
The MK formula ignores the possible polarization
dependence, because it deals with
$\chi''_{+-}\sim \chi''_{xx}+ \chi''_{yy}$, and
not $\chi''_{xx}$ and $\chi''_{yy}$ separately.
The polarization dependence was discussed theoretically first
by Natsume {\it et al.}~\cite{Natsume1,Natsume2,Natsume3}
generalizing the Kubo-Tomita theory.
It has also been observed experimentally~\cite{Natsume1,Natsume2}
and numerically~\cite{Miya-ESR}.

However, apparently it has been not recognized that,
for some special cases,
an {\em exact and rigorous} result on the polarization dependence
can be derived easily from the equation of motion.
Let us consider the special case in which the perturbation
$\calH'$ is written in terms of the $x$ component of the
spin operator $S^x_j$.
The examples include the transverse
staggered field $\calH' = h \sum_j (-1)^j S^x_j$,
and the exchange anisotropy with the anisotropy
axis in the $x$ direction
$\calH' = \delta \sum_j S^x_j S^x_{j+1}$.
In these cases, $[S^x, \calH' ] =0$ holds, and consequently
\begin{equation}
    \frac{d S^x}{dt} = H S^y .
\end{equation}
This identity leads to
\begin{equation}
\chi''_{xx}(0,\omega) = \frac{H^2}{\omega^2} \chi''_{yy}(0,\omega) ,
\label{eq:pol-identity}
\end{equation}
and more generally, for the polarization in the direction
$\alpha$ in the $xy$-plane,
\begin{equation}
\chi^{\alpha \alpha}(0,\omega) =
 \frac{H^2\cos^2{\Phi}+\omega^2\sin^2{\Phi}}{\omega^2} \chi^{yy}(0,\omega) ,
\label{eq:pol-dependence}
\end{equation}
where $\Phi$ is the angle between $x$ and $\alpha$.
(In the notation of Refs.~\onlinecite{Natsume1,Natsume2,Natsume3},
$\theta=90$ degree and their $\phi$ corresponds to our $\Phi$.)

For a sharp resonance concentrated near $\omega \sim H$,
the polarization dependence is not significant.
However, if the center of the resonance is defined by
the average frequency of the spectrum
\begin{equation}
\bar{\omega}_{\alpha} =
\frac{\int \omega I_{\alpha} (\omega) d \omega }{
        \int I_{\alpha} (\omega)d \omega},
\end{equation}
$\bar{\omega}_x < \bar{\omega}_y$ because the higher frequency
part is emphasized in the latter.
As a consequence, there is a positive frequency shift
for the polarization in $y$ axis, compared to the case where
in $x$ axis.

This is in agreement with theoretical and experimental results
in Refs.~\onlinecite{Natsume1,Natsume2} and numerical results
in Ref.~\onlinecite{Miya-ESR}, on the exchange anisotropy.
We note that, in the actual experiments on ESR, the resonance
frequency is kept fixed and the applied field is scanned to
measure the absorption.
Because of this, it is
customary to  discuss the shift of the resonance field for a
fixed frequency. The direction (positive or negative)
of the field shift is opposite to that of the frequency shift
we discuss in this paper.
They find that the resonance field is shifted negatively
for the polarization in $y$ direction compared
to the $x$ polarization case,
which is indeed consistent with our result.
The angular dependence is also consistent with
the theoretical formula in Ref.~\onlinecite{Natsume1}.
On the other hand, in Ref.~\onlinecite{Natsume3},
the polarization dependence is studied by a different
formalism (Mori's memory function method.)
When the anisotropy axis is perpendicular to the applied field,
the obtained polarization dependence is rather opposite
to the above, and is in contradiction to our
rigorous result~(\ref{eq:pol-identity}).

\section{Field-theory approach to the $S=1/2$ Heisenberg
antiferromagnetic chain}
\label{sec:ft}

\subsection{Bosonization of $S=1/2$ Heisenberg chain}
\label{sec:bosonization}

In the present paper, we mainly discuss ESR on the one-dimensional $S=1/2$
Heisenberg antiferromagnet
\begin{equation}
\calH_0 = J \sum_j \vec{S}_j \cdot \vec{S}_{j+1} .
\label{eq:HAFC}
\end{equation}
with symmetry-breaking perturbation $\calH'$
and of course the Zeeman term $\calH_Z$.
The low-energy physics of the one-dimensional quantum antiferromagnets
is well described by field theory methods (bosonization).
In this section, we briefly summarize the aspects of this approach
that are relevant to the present discussion of the ESR.
We refer the reader to Refs.~\onlinecite{Affleck:LesHouches,GNTbook}
for more details.
While the method is now standard,
here we also clarify subtleties specific
to ESR problems, which are related to the
weakly broken $SU(2)$ symmetry discussed in Section~\ref{sec:esr}.

The effective field theory of the $S=1/2$ Heisenberg chain $\calH_0$
is given by the free boson Lagrangian
\begin{equation}
{\cal L} =
\frac{1}{2} \left[ ( \partial_0 \phi )^2 - ( \partial_1 \phi )^2 \right],
\label{eq:freebosonL}
\end{equation}
where $x^0 = vt$, $x^1 = x$ and we make identification
$\phi \sim \phi + 2 \pi R$ with the compactification radius $R$.
The radius $R$ is actually fixed to the value $1/\sqrt{2 \pi}$
by the $SU(2)$ symmetry.
Hereafter we set $v = 1$ for simplicity; the spinon velocity $v$
can be recovered by dimensional analysis when necessary.

At zero uniform field, the spin operators
may be written in terms of the field $\phi$ as follows:
\begin{eqnarray}
    S^z_j &\sim& \frac{1}{ 2 \pi R} \frac{\partial \phi}{\partial x}
        + C_s^z (-1)^j \cos{\frac{\phi}{R}}
\\
    S^-_j &\sim&
        C_u^-  e^{- i 2 \pi R \tilde{\phi}}
                \cos{\frac{\phi}{R}}
            + C^-_s (-1)^j e^{- i 2 \pi R \tilde{\phi}},
\end{eqnarray}
where the dual field $\tilde{\phi}$ is defined in terms of
right-mover $\varphi_R$ and $\varphi_L$ as
$\phi = \varphi_R + \varphi_L$ and $\tilde{\phi} = \varphi_R - \varphi_L$.
While $S^z$ and $S^{x,y}$ are represented in a very
different way, their correlation functions turn out to be equal
at the $SU(2)$ invariant radius $R=1/\sqrt{2\pi}$, as required
from the symmetry of the original Heisenberg chain.

\begin{figure}
\begin{center}
\epsfxsize=8cm
\epsfbox{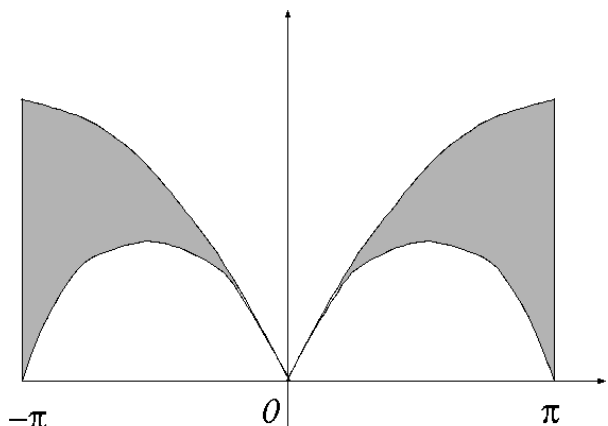}
\caption{
The spin structure factor of the $S=1/2$ Heisenberg antiferromagnetic
chain at $T=0$.
It is non-vanishing only in the filled region
shown in the frequency-momentum plane.
The structure factor becomes a delta function $\delta(\omega-vq)$
in the $q\rightarrow 0$ limit.
}
\label{fig:struct}
\end{center}
\end{figure}

The dynamical structure factor ${\cal S}^{\alpha \alpha}$
(Fourier transformation of the spin correlation function)
of the Heisenberg chain has been studied in detail.
It is equivalent
to the dynamical susceptibility for $T=0$ and $\omega>0$.
At zero temperature, the dynamical structure factor
is non-vanishing only in the limited
region of the frequency $\omega$ -- momentum $q$ space shown
in Fig.~\ref{fig:struct}.
The field theory actually can handle only the low-energy
excitations near momentum $0$ and $\pi$.
The structure factor for ${\cal S}^{zz}$ near $q=0$ and $q=\pi$ is given by
the correlation function of $\partial \phi / \partial x$
and $\cos{(\phi/R)}$ respectively.
At $T=0$, they read
\begin{equation}
    {\cal S}^{zz} (\omega,q)\propto \delta(\omega - |q|) \theta(\omega)
\end{equation}
for $q \sim 0$ and
\begin{equation}
{\cal S}^{zz} (\omega,q) \propto
\frac{1}{\sqrt{\omega^2 - (q-\pi)^2}} \theta(\omega -|q-\pi|)
\end{equation}
for $q \sim \pi$.
It is noted that the structure factor is completely sharp
and is delta-function like at $q \sim 0$.
In fact, the structure factor at $q \sim 0$
remains so even at finite temperature.
As mentioned before, the structure factor is of course isotropic
(${\cal S}^{xx} = {\cal S}^{yy} = {\cal S}^{zz}$)
at $H=0$ for the isotropic Heisenberg chain.

Now let us consider the effect of the applied magnetic field.
The Zeeman term $\calH_Z$ in the Lagrangian becomes, upon
bosonization,
\begin{equation}
{\cal L}_H={H\over \sqrt{2\pi}}{\partial \phi \over \partial x}.
\end{equation}
This term can be eliminated by a redefinition of the boson field
\begin{equation}
\phi (t,x)\to \phi (t,x)+{H\over \sqrt{2\pi}}x.
\label{eq:phishift}
\end{equation}
but $\tilde{\phi}$ remains unchanged.
This is equivalent to the shift of chiral fields as
\begin{equation}
    \varphi_R \rightarrow \varphi_R + \frac{1}{2 \sqrt{2 \pi}} H x ,
    \varphi_L \rightarrow \varphi_L + \frac{1}{2 \sqrt{2 \pi}} H x .
\label{eq:phiRLshift}
\end{equation}
While this leaves the free Lagrangian unchanged, it does change
the bosonization formulae of physical spin operators:
\begin{eqnarray}
    S^z &\sim& m + \frac{1}{ 2 \pi R} \frac{\partial \phi}{\partial x}
            + C_{s}^z \cos{[ \frac{\phi}{R} + (H + \pi) x]},
\label{eq:Szops}
\\
    S^{\pm} &\sim&
        C_u^-  e^{- i \sqrt{2 \pi} \tilde{\phi}}
                \cos{( \frac{\phi}{R} + H x)}
            + C^-_s (-1)^j e^{- i 2 \pi R \tilde{\phi}}.
\label{eq:S-ops}
\end{eqnarray}
The first term $m$ in $S^z$ represents the expectation value of the
magnetization induced by the magnetic field $H$.
For a small magnetic field, $m$ is proportional to the field $H$.
Another important feature is that
the applied field induces
the shift of the soft-mode momentum\cite{IshimuraShiba,Mueller}.
The shift occur differently for the longitudinal ($z$) and
the transverse ($x,y$) components.
The gapless points under the applied uniform field $H$ are
at $q=0$ (uniform part) and $q=\pi \pm H$ (``staggered'' part)
for the longitudinal modes.
For the transverse modes,
they are at $q=\pm H$ (``uniform'' part) and $q=\pi$ (staggered part.)

\begin{figure}
\begin{center}
\epsfxsize=8cm
\epsfbox{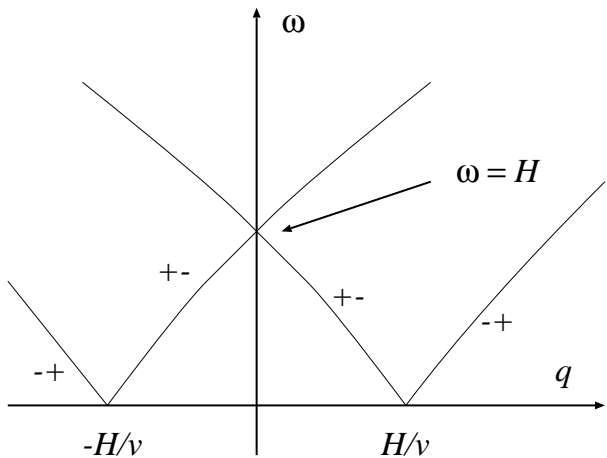}
\caption{
The zero temperature transverse spin structure
factor $S_{xx}(\omega,q)=S_{yy}(\omega,q)$
of the $S=1/2$ Heisenberg antiferromagnetic chain under an
applied field $H$, near $q =0$.
It is approximately proportional to
$\omega [ \delta(\omega - |q -H|) + \delta(\omega - |q+H|)]$,
giving the resonance at $q=0, \omega = H$.
This consists of two branches coming from $S_{+-}$ and $S_{-+}$,
which are marked by $+-$ and $-+$ in the graph.
In fact, there is a small spreading of the spectrum and
the structure factor is generally not a perfect delta function.
However, it is exactly the delta function $\delta(\omega -H)$
at $q=0$, as explained in the text.
}
\label{fig:transH}
\end{center}
\end{figure}

Let us focus on the transverse
mode near $q=0$, because
the transverse mode at $q=0$ is measured in ESR
in the Faraday configuration.
For simplicity, here we restrict ourselves to zero temperature.
In the low energy effective theory,
the ``uniform'' part of the $S^{\pm}$ is given
\begin{equation}
S^{\pm} \propto e^{\pm (iHx+i\sqrt{8\pi}\phi_R)}+
 e^{\mp (iHx + i\sqrt{8\pi}\phi_L)} ,
\end{equation}
where we have used the $SU(2)$ symmetric compactification
radius $R=1/\sqrt{2\pi}$ (see below for reason for taking this value.)
This gives the correlation function of $S^{\pm}$ at zero temperature: 
\begin{equation}
\langle S^+(t,x)S^-(0,0) \rangle \propto {e^{iHx}\over (t-i\epsilon +x)^2}
+ {e^{-iHx}\over (t-i\epsilon -x)^2}
\end{equation}
Dynamical structure factor $S_{+-}$, which is the Fourier transform
of the above is,
\begin{eqnarray}
S_{+-}(\omega,q ) &\propto& \int_{-\infty}^\infty dt\int_{-\infty}^\infty dx
e^{i(\omega t-qx)}\left[ {e^{iHx}\over (t-i\epsilon +x)^2}
+ {e^{-iHx}\over (t-i\epsilon -x)^2}\right]\nonumber \\
&=& \int_{-\infty}^\infty dx[-2\pi \omega \theta(\omega )
 ]e^{-iqx} \left[
e^{iHx-i\omega x}+e^{-iHx+i\omega x}\right]\nonumber \\
&\propto&
\omega \theta(\omega ) [\delta (\omega -H+q)+
\delta (\omega -H-q)] .
\label{eq:S+-}
\end{eqnarray}
The other one $S_{-+}$ is given by replacing $H \rightarrow -H$
in the above, using the time reversal transformation.
Thus
\begin{equation}
S_{-+}(\omega,q ) \propto
\omega \theta(\omega ) [\delta (\omega + H+q)+\delta (\omega + H-q)] .
\label{eq:S-+}
\end{equation}
Namely, $S_{+-}$ and $S_{-+}$ give different branches of excitation.
The fact that $S_{-+}$ does not contain
the branch~(\ref{eq:S+-}) was recognized earlier
(see Fig.~17 of Ref.~\onlinecite{Mueller}.)
On the other hand, that $S_{+-}$
lacks the branch~(\ref{eq:S-+}) (at least in the low-energy limit)
was apparently not appreciated in
Fig.~18 of Ref.~\onlinecite{Mueller}.
$S_{xx}$ and $S_{yy}$ are given by their superposition
\begin{equation}
S_{xx}(\omega,q ) = S_{yy}(\omega,q) \propto
\omega [\delta (\omega - |q+H|)+\delta (\omega - |q - H|) ] .
\end{equation}
This zero-temperature transverse structure factor near $q=0$ under the
applied magnetic field is shown in Fig.~\ref{fig:transH}.
Because the structure factor near $q=0$ was sharp, and
the gapless point is shifted by $H$, we expect a sharp
resonance at energy $\omega \sim H$ at $q=0$.
This corresponds to the expected paramagnetic ESR for the
isotropic Heisenberg chain.

However, it should be noted that we have so far ignored various
renormalization effects due to the applied magnetic field.
There are irrelevant operators, which themselves vanish
in the low-energy limit
but renormalize parameters of the low-energy effective theory.
The way they renormalize is affected by the applied magnetic field.
In general, the precise value of
the momentum shift is given by $2 \pi m$ rather than $H$,
where $m$ is the magnetization.
This can be derived from the shift of Fermi momentum in the Jordan-Wigner
transformation, and also is required from a rigorous version of
Luttinger's theorem in one dimension\cite{YOA}.
Restoring the spinon velocity $v$, the ESR frequency appears to be given
by $2 \pi m v$.
For the standard Heisenberg antiferromagnetic chain in an applied field,
the magnetization $m$ and the spinon velocity $v$
can be obtained as a function of $H$
from the Bethe Ansatz integral equation.
Generally, $2 \pi m v$ is different from $H$ except in the
zero field limit,
implying that the ESR frequency deviates from $H$.
However, this cannot be true, because the equation of motion
for the original Heisenberg model (under an applied field) requires
the resonance to be exactly at the frequency $H$.
The resolution is that, the dispersion relation for $q \sim 0$ is
not completely linear.
The curvature of the dispersion comes form irrelevant operators
which break Lorentz invariance.
Because of the curvature,
the resonance frequency at $q=0$ is modified from $2 \pi m v$, which
is derived assuming the linear dispersion.
What the equation of motion tells us is that these renormalization
effects miraculously cancel, to give the resonance exactly
at $\omega = H$ for $q=0$.
With this nontrivial mechanism in mind, we will take the momentum
shift as $H$, setting the spinon velocity $v=1$.

There is another ``miraculous'' cancellation similar
to the above.
At zero field, the compactification radius of the effective
field theory is fixed to the special $SU(2)$ symmetric value
$R=1/\sqrt{2\pi}$,
as is required from the $SU(2)$ symmetry of the original
Heisenberg model.
However, in the presence of the applied field, the $SU(2)$
symmetry is of course broken down to $U(1)$.
Correspondingly, the radius $R$ is renormalized away from
the $SU(2)$ point by the applied field.
The renormalized radius $R$ as a function of the applied
field $H$ has been also obtained from the exact Bethe Ansatz
solution~\cite{BIK}. It is indeed rather sensitive
to $H$ for small $H/J$.
A consequence of the radius renormalization is the
dependence of the correlation exponents on the applied field.
In particular, the ``uniform'' part of the transverse spin
operator, which is relevant for ESR,
is represented by the vertex operator of the type
$\exp{[ \pm 2 \pi i R \tilde{\phi} \pm \phi/R]}$;
its conformal weight is given by
\begin{equation}
(\Delta, \bar{\Delta}) = ( 1 + \Delta', \Delta')
\label{eq:Delta}
\end{equation}
or $ ( \Delta', 1 + \Delta')$,
where
\begin{equation}
 \Delta' = \frac{(2 \pi R - 1/R)^2}{8 \pi},
    \label{eq:Delta-R}
\end{equation}
which does depend on $R$.
As a result, the structure factor is no longer given by
a delta-function for $R \neq 1/\sqrt{2 \pi}$.
More explicitly, the retarded Green's function of a conformal primary field
with conformal weight $(\Delta,\bar{\Delta})$
at finite temperature $T$ is obtained explicitly~\cite{Schulz} as
\begin{equation}
G^R_{(\Delta,\bar{\Delta})} (\omega, q) =
- \sin{(2 \pi \Delta)} ( 2 \pi T)^{2 (\Delta + \bar{\Delta} -1)}
B(\Delta - i \frac{\omega + q}{4 \pi T}, 1 - 2 \Delta)
B(\bar{\Delta} - i \frac{\omega -  q}{4 \pi T}, 1 - 2 \bar{\Delta}) ,
\label{eq:Schulz}
\end{equation}
where $B$ denotes the Euler Beta function:
\begin{equation}
B(x,y)={\Gamma (x)\Gamma (y)\over \Gamma (x+y)}\end{equation}
and $\Gamma$ is Euler's Gamma function.
Considering the momentum shift induced by the applied field,
the absorption measured in ESR corresponds to the
Green's function evaluated at $q=H$.
Thus, the spectrum is given by the delta function only if
$(\Delta, \bar{\Delta}) = (1,0)$ or $(0,1)$, namely $R=1/\sqrt{2 \pi}$.
The renormalization of $R$ due to the applied field seems to imply that
the ESR spectrum should not be given by a delta-function,
even in the absence of the perturbation $\calH'$.

However, this is inconsistent with
the equation of motion of the original Heisenberg model.
It predicts a completely sharp (delta-function) resonance precisely at
the Zeeman energy even for a finite field $H$.
Since the equation of motion is exact and rigorous for
the original spin problem, we conclude that we should
take the unrenormalized, $SU(2)$ symmetric value
$R=1/\sqrt{2\pi}$ even in a finite field,
for the calculation of the ESR.
This appears contradictory to the well-established
renormalization of $R$ due to the applied field.
This is not a real contradiction, however, because the standard
result on the renormalization of the radius is determined at
the zero energy limit, while the ESR probes the excitation at
the finite energy $H$.
In general, effective coupling constants depend on the
energy scale as a consequence of the renormalization.
We may introduce an effective radius $R(\omega)$ as a function
of the energy scale $\omega$.
While the determination of the function $R(\omega)$ in general
is a tedious task, the exact equation of motion on ESR gives
the restriction at the Zeeman energy: $R(\omega = H) = 
1/ \sqrt{2\pi}$.
The non-renormalization could be related to the qualitative understanding
of the RG flow in the presence of the applied field, Fig.~7 in
Ref.~\onlinecite{Cubenz-PRB}. 
The RG flow in the presence of the applied field is
almost identical to that in the zero field, down to energy scale of $O(H)$,
where the flow is ``cut off.''
If we look at the energy $H$, the effective theory may be almost identical
to the isotropic one.
This argument would not, however, explain why the effective
radius should be exactly at the $SU(2)$ point.
{F}rom the viewpoint of the field theory, this is again
a miraculous cancellation between the renormalization
by the uniform field and that by the finite energy.
The equation of motion, although quite simple,
gives an exact and highly nontrivial constraint on the effective
field theory description.

Thus, in the following calculations we do not include the
radius renormalization due to the applied field,
and take the $SU(2)$-symmetric value $R=1/\sqrt{2 \pi}$.
As a result, the appropriate effective field theory of ESR is
an $SU(2)$ symmetric one, namely the level-1 $SU(2)$ Wess-Zumino-Witten
(WZW) theory, even in a finite field;
all the effects of the applied field are represented by
the shift of the $\phi$ field~(\ref{eq:phishift}),
resulting in the momentum shift~(\ref{eq:Szops}) and (\ref{eq:S-ops}).
This may be regarded as a field theory representation
of the crucial $SU(2)$ symmetry which is broken only weakly,
discussed in Section~\ref{sec:esr}.

It is often convenient to introduce the operators in
non-Abelian bosonization to make the symmetry manifest.
$SU(2)$ current operators $J^{\alpha}$ ($\alpha = x,y,z$)
are related to the Abelian bosonization as follows:
\begin{eqnarray}
J^z_R (w)  &=& i \sqrt{4 \pi} \partial \varphi_R(w) ,
\label{eq:JzRphi}
\\
J^{\pm}_R(w) &=&
     \sqrt{2} e^{\pm i \sqrt{8 \pi} \varphi_R (w)},
\\
J^z_L (\bar{w})  &=&
    - i \sqrt{4 \pi} \bar{\partial} \varphi_L(\bar{w}),
\label{eq:JzLphi}
\\
J^{\pm}_L(\bar{w}) &=&
    \sqrt{2} e^{\mp i \sqrt{8 \pi} \varphi_L (\bar{w})},
\end{eqnarray}
where $J^{\pm} = J^x \pm i J^y$,
we have introduced complex coordinates $w = \tau + i x$
($\tau = it$) and $\phi(w,\bar{w}) = \varphi(w) + \bar{\varphi}(\bar{w})$.
$J^{\alpha}_{R(L)}$ is the right-mover (left-mover)
component of the current, and we have normalized them by
\begin{equation}
    \langle J^a_R(w_1) J^b_R(w_2) \rangle
            = \frac{\delta^{ab}}{(w_1 - w_2 )^2},
\label{eq:JJcorr}
\end{equation}
where $a,b = x, y, z$ and the complex coordinate
$w = \tau + i x = - i (t - x)$ and likewise for the $L$ sector.
(We note that this is different normalization from
Ref.~\onlinecite{AGSZ}.)

The ``uniform'' part of the spin operators $S^{\alpha}$ correspond to
the $SU(2)$ currents $J^{\alpha}$, while the ``staggered'' part is related to
the $SU(2)$ triplet $n^{\alpha} = {\rm Tr}{g \sigma^{\alpha}}$
where the $SU(2)$ matrix field $g^\alpha_\beta (x,t)$ is the
fundamental field of the Wess-Zumino-Witten non-linear $\sigma$-model.
Eqs.~(\ref{eq:Szops}) and (\ref{eq:S-ops}) may be rewritten as
\begin{eqnarray}
S^z &\sim& \frac{1}{\sqrt{8 \pi^2}} ( J^z_R + J^z_L )
    + C_s \left[ \cos{(H+\pi) x} n^z + \sin{(H+\pi)x} {\rm tr}g \right],
\label{eq:SzopsN}
\\
S^{\pm} &\sim&
\frac{1}{\sqrt{8 \pi^2}} ( J^{\pm}_R e^{\pm i Hx} + J^-_L e^{\mp i Hx}) .
+ (-1)^x C_s n^{\pm},
\label{eq:S-opsN}
\end{eqnarray}
The ``staggered'' part of $S^z$ may be written as $(-1)^x n^z$ at
$H=0$, but is a mixture of $n^z$ and ${\rm tr}g$ in a finite field.

The ESR absorption intensity is related to the Green's function
of $S^{x,y}$; thus what is needed in the field theory
is the Green's function of $J^{x,y}$ at momentum $\pm H$.

\subsection{Perturbations}
\label{sec:perturbations}

Having established the effective field theory for
the unperturbed system $\calH_0 + \calH_Z$,
we now want to calculate the effects of
the perturbation $\calH'$ on the ESR lineshape.
Assuming that the perturbation $\calH'$ is small,
$\calH'$ can be mapped to an operator of the level-1 $SU(2)$ WZW theory.

In principle, an infinite variety of symmetry breaking
perturbations $\calH'$ is possible. In fact, there are infinitely
many operators also in the field theory.
However, most of the operators have large scaling dimensions, and
thus renormalize rapidly to zero under the RG transformation.
Thus, at low enough temperatures, only a few types of perturbations
with smaller scaling dimensions are important.

The operators with the lowest scaling dimension $1/2$ are
$n^{\alpha}$ and ${\rm tr}g$ in WZW theory.
In the original spin chain Hamiltonian (at $H=0$), they correspond to
the staggered field (3 independent perturbations corresponding
to three directions) and the bond alternation.
However, the bond-alternation does not break the $SU(2)$
symmetry and hence should not affect the ESR lineshape, although
it is not trivial to see this in the field theory.
On the other hand, the staggered field perturbation does
break the $SU(2)$ symmetry and thus affects the ESR lineshape.
The operators of interest with the second lowest scaling dimension $2$,
which are marginal,
are $J^{\alpha}_L J^{\alpha}_R$.
They correspond to the exchange anisotropy in the spin chain
Hamiltonian.
We will discuss these two most important cases in later sections.

While we use the $SU(2)$ symmetric field theory, care should
be taken with the momentum shift due to
the applied field.
The momentum shift is determined by a simple rule in Abelian
bosonization formulation~(\ref{eq:phiRLshift}).
Namely, if one writes some operator at zero field
in terms of  $\varphi$'s, the above replacement gives
a correct formula under the finite field $H$.
The operator corresponding to the
perturbation $\calH'$ may contain an oscillating
factor. While such a term may be ignored in order to know
whether there is a finite excitation gap above the ground state,
it should be retained in theory of ESR which probes finite momentum
of the effective field theory.
For a general perturbation,
the oscillating factor appears in the effective field theory,
and it makes the theoretical analysis rather complicated.
In this paper, we focus on a few simple cases in which there is
no oscillating term (with finite momentum) in the effective Lagrangian.
This still includes several cases of physical interest
which are mentioned below.

\subsubsection{Transverse staggered field}
\label{sec:stagfield}

A quasi one-dimensional spin system often has an alternating
crystal structure along the chain.
In such a case, generally we expect two features which are absent
in a uniform system.
\begin{description}
\item[staggered $g$-tensor]
The magnetic field $\vec{H}$ couples to the spin as
$\mu_B\sum_{j,a,b}H^a [g^u_{ab}+(-1)^jg^s_{ab}]S^b_j$,
where $g^s$ is the staggered component of the $g$-tensor.
\item[Dzyaloshinskii-Moriya (DM) interaction]
The low symmetry allows the antisymmetric interaction\cite{Dzyal,Moriya}
   $\sum_j \vec{D}_j \cdot (\vec{S}_j \times \vec{S}_{j+1} )$ .
\end{description}
The DM interaction can be either uniform ($\vec{D}_j = \vec{D}$)
or staggered ($\vec{D}_j = (-1)^j \vec{D}$.)

When the staggered $g$-tensor is present, an effective
staggered field $\propto g^s \vec{H}$ is produced
upon an application of the external field.
The direction of the staggered field is often approximately
perpendicular to the applied field, although it is not necessarily so.
The effect of the DM interaction is less trivial, but it can be
actually eliminated by an exact transformation.
Let us consider the case of a staggered DM interaction,
and choose the axes so that the DM vector $\vec{D}$ is parallel
to the $z$-axis.
Then the Hamiltonian including the DM interaction is given by
\begin{eqnarray}
  \calH_{DM} &=&
J \sum_j \vec{S}_j \cdot \vec{S}_{j+1} +
(-1)^j D (S^x_j S^y_{j+1} - S^y_j S^x_{j+1} )
\nonumber \\
&=&
\frac{1}{2} \sum_j [ {\cal J} S^+_{2j-1} S^-_{2j}
                          + {\cal J}^*  S^+_{2j} S^-_{2j+1}
                          + (\mbox{h.c.}) ]
\nonumber \\
&&     + J \sum_j [ S^z_{2j-1} S^z_{2j}+S^z_{2j}S^z_{2j+1}],
\end{eqnarray}
where ${\cal J} \equiv J+iD$.
Now let us define the angle $\alpha = \tan^{-1}{D/J}$,
and rotate the spin at site $j$ by the angle $(-1)^j \alpha/2$
about $z$  axis:
\begin{equation}
 S^+_j \rightarrow S^+_j e^{i (-1)^j \alpha/2}
\label{eq:DMelim}
\end{equation}
Then we obtain the Hamiltonian of the XXZ chain
\begin{equation}
 \hat H=\sum_j[J S_j^z S_{j+1}^z +
     \frac{|{\cal J}|}{2}( S_j^+ S_{j+1}^- + h.c.)].
\end{equation}
It is argued\cite{KSEA} that
this anisotropic exchange can cancel the pre-existing one.

Now suppose that an external field $H$ is applied in $x$ direction.
The applied field is transformed as
\begin{equation}
-H\sum_j S^x_j \to
-H\sum_j
 [\cos{\frac{\alpha}{2}} S^x_j+(-1)^j \sin{\frac{\alpha}{2}} S^y_j].
\end{equation}
by the above transformation.
Thus, in the presence of the Dzyaloshinskii-Moriya interaction,
the applied uniform field
produces an effective staggered field\cite{Cubenz-PRL}.
For general orientations of $\vec{D}$ of the staggered DM interaction
\begin{equation}
\calH_{DM} = \sum_j (-1)^j \vec{D} \cdot (\vec{S}_j \times \vec{S}_{j+1}) ,
\label{eq:DMterm}
\end{equation}
the effective staggered field due to the DM interaction is
given by $\vec{D}\times\vec{H}/(2J)$

These two effects give an effective transverse staggered field
which is approximately perpendicular to the applied field.
This mechanism is important in studying properties of
several quasi-one dimensional antiferromagnets
including Cu benzoate\cite{Dender1,Cubenz-PRL,Cubenz-PRB},
Yb$_4$As$_3$\cite{Yb4As3} and Pyrimidine Cu dinitrate\cite{Feyerherm}.

\subsubsection{Exchange Anisotropy}

The exchange anisotropy is the second relevant perturbation
which affects the ESR lineshape.
The dipolar interaction which exists in any real magnetic
system is given by, restoring the Bohr magneton $\mu_B$,
\begin{equation}
 H_{dp} = (g \mu_B)^2 \sum_{ij} \left[
    \frac{ \vec{S}_i \cdot \vec{S}_j}{|\vec{r}_{ij}|^3}
    - \frac{3 (\vec{S}_i \cdot \vec{r}_{ij})
        (\vec{S}_j \cdot \vec{r}_{ij}) }{|\vec{r}_{ij}|^5}
\right],
\end{equation}
where $\vec{r}_{ij}$ represents the vector from site $i$ and $j$
and for the simplicity
the $g$-factor is assumed to be uniform and isotropic.
In a spin chain, the vector $\vec{r}_{ij}$ is parallel to the
chain direction, and the dipolar interaction reduces to an
effective exchange anisotropy parallel to the chain direction. The
effect would be essentially the same with the nearest-neighbor
anisotropic exchange interaction, because the dipolar interaction
strength decreases rapidly with the distance.

Let us consider the simplest case of the exchange anisotropy
\begin{equation}
 H_a = \delta \sum_j S^n_j S^n_{j+1}
\label{eq:anis}
\end{equation}
with a symmetry axis $n$, which effectively covers the case of the
dipolar interaction if $n$ is taken to be the chain direction.
Even in this simple case, a variety of configurations is possible
by changing the relative direction of $n$ and the direction $z$ of
the applied field, as is often done in experiments.

As mentioned before, for a general direction, the perturbation
in the field theory is rather complicated, making a calculation
from first principles difficult.
Thus, in this paper, we will focus on the two simplest cases, namely
when $n \parallel z$ and $n \perp z$.
The case $n \parallel z$ allows us a direct calculation of the
lineshape and will be discussed in Section~\ref{sec:anis-z}.
The latter case $n \perp z$ will be discussed in
Section~\ref{sec:anis-x}, based on the self-energy
approach developed in Section~\ref{sec:self-energy}.

\section{Exchange anisotropy parallel to the field: direct calculation}
\label{sec:anis-z}

Here we consider the case where the anisotropy axis is parallel
to the applied magnetic field, namely $n=z$ in (\ref{eq:anis}).
In this case, it is obvious that there is no polarization
dependence as $S^x$ and $S^y$ are equivalent.

In this case, the perturbation in the effective field theory
is given, {\em at zero magnetic field}, as
\begin{equation}
 {\cal L}_a = - \lambda J^z_R J^z_L ,
\label{eq:anisz-eff}
\end{equation}
where $\lambda$ is a parameter proportional to $\delta/J$,
for a small anisotropy $\delta/J$.
The proportionality constant $\lambda J / \delta$ is non-universal
and model-dependent.
(For the standard Heisenberg antiferromagnetic chain, $\lambda$
is determined in Section~\ref{sec:logcorr} together with a logarithmic
correction.)

Before performing an explicit calculation, let us see what can be
said about the temperature dependence of the linewidth from a general
scaling argument. The perturbation~(\ref{eq:anisz-eff}) is a
marginal one with the scaling dimension $2$. Thus, ignoring the
logarithmic corrections,  scaling arguments imply that the
linewidth takes the scaling form
\begin{equation}
    \eta = T f(\frac{\delta}{J}, \frac{H}{T}),
\label{eq:deltascale}
\end{equation}
where we have used the fact that $\eta$ has the dimension of
 energy. In fact,  this scaling argument can be applied to any
direction of the applied field. On the other hand, the explicit
form of the scaling function $f$ cannot be determined by the
scaling argument alone.

Now let us calculate the linewidth explicitly for the
anisotropy parallel to the applied field.
As we have discussed, all the effect of the applied uniform field
is represented by the shift of the $\phi$ field~(\ref{eq:phishift}).
Consequently, the perturbation under the applied field $H$ is
\begin{equation}
 {\cal L}_a =
- \lambda J^z_R J^z_L  - \frac{\lambda H}{\sqrt{2}}( J^z_R + J^z_L )
- \frac{\lambda H^2}{2} .
\label{eq:aniszLag}
\end{equation}
The third term is a constant and thus can be ignored.
The second term is
\begin{equation}
- \frac{\lambda H}{\sqrt{2}}( J^z_R + J^z_L )
 = - 2 \pi \lambda H \frac{1}{\sqrt{2\pi}} \frac{\partial \phi}{\partial x},
\end{equation}
which is equivalent to the additional magnetic field of
$- 2 \pi \lambda H$.
This can be absorbed by a renormalization of the magnetic field,
giving  the shift of the resonance by $ - 2\pi \lambda H$.
This shift is first order in the perturbation $\delta$ and the field $H$.

Now, let us discuss the effect of the first term.
We should calculate the correlation function
$\langle J^+ J^- \rangle$ in the presence of the
perturbation $- \lambda J^z_R J^z_L$.
For this particular problem,
this can be done exactly, because
the perturbation $J^z_R J^z_L$ is proportional to the
kinetic term of the free boson Lagrangian; it just gives
a renormalization of the compactification radius.
That is, the Lagrangian density reads
\begin{eqnarray}
{\cal L} &=&
\frac{1}{2}(\partial_{\mu} \phi)^2 - \lambda J^z_L J^z_R
\nonumber \\
&=& \frac{1 + 2 \pi \lambda}{2} (\partial_{\mu} \phi)^2 .
\end{eqnarray}
Rescaling the field $\phi$ so that the coefficient of the
kinetic term is again given by $1/2$,
the renormalized radius $R$ is given as
\begin{equation}
    R = \sqrt{\frac{1 + 2 \pi \lambda}{2\pi}} .
\end{equation}
We note that, we have not included the similar renormalization due
to the applied field because of the subtleties explained in
Section~\ref{sec:bosonization}. In contrast, the exchange
anisotropy does break the $SU(2)$ symmetry; there is no reason not
to include the renormalization in the present case.

The conformal weight of the vertex operator
$J^{\pm} = e^{\pm i 2\pi R \tilde{\phi} + i \phi/R}$ is
$(\Delta, \bar{\Delta}) = ( 1 + \Delta', \Delta')$ or
$ ( \Delta', 1 + \Delta')$ where
$\Delta' = ( 2 \pi R - 1/R)^2/(8 \pi) \sim \pi^2 \lambda^2$.
Its Green's function at finite temperature is given
in eq.~(\ref{eq:Schulz}).
As explained, the Green's function evaluated at
the momenta $\pm H$ is relevant for ESR.
Near the center of the resonance, the spectrum is dominated by
the pole of the $\Gamma$ function;
it reduces to
\begin{equation}
{\calG^R}_{S^+ S^-} (\omega) \sim
\frac{\mbox{const.}}{\omega - H + 4 \pi T \Delta' i} .
\label{eq:Gamma-Lorentz}
\end{equation}
Thus the resonance is Lorentzian with the width
\begin{equation}
    \eta = 4\pi \Delta' T = 4 \pi^3 \lambda^2 T .
\label{eq:zzwidth}
\end{equation}
This is  consistent with the  scaling
argument~(\ref{eq:deltascale}).

To summarize, the exchange anisotropy with the axis
parallel to the applied field gives the following effects
on paramagnetic ESR.
\begin{description}
\item[shift] $-  2 \pi \lambda H \propto - H \delta$
\item[width] $4 \pi^3 \lambda^2 T \propto (\delta/J)^2 T$
\end{description}

\section{Self-energy approach}
\label{sec:self-energy}

In the last Section, the ESR absorption spectrum was calculated
directly in the low-energy effective theory. This was made possible
because the effective theory reduced to the free boson theory.
However, in general, the problem is more difficult because the
effective field theory involves interactions.

A possible application of the field theory method to ESR is to evaluate the
Green's function appearing in MK formula~(\ref{eq:MKwidth})
by means of the field theory.
While the MK formula has been applied to quantum spin chains
by several authors, most of the calculations are based
on classical or high-temperature approximations which
break down at low temperature and in low dimensions.
Thus it would be worthwhile to evaluate the MK formula
using field theory to study quantum spin systems at lower
temperature and in lower dimensions.
On the other hand, the crucial assumption of the
(single) Lorentzian lineshape is made in using the MK
formula usually without a rigorous justification.
Moreover, the MK formula ignores the possible polarization
dependence discussed in Sec.~\ref{sec:polari}.
Thus, in this section, we develop a new, systematic
field-theory approach to ESR, which we call the self-energy approach.
The ESR spectrum  is given by the imaginary part
of the retarded Green's function of $S^{\pm}$.
As we have discussed in the last section, it corresponds
to the Green's function of the current operators in the
effective field theory via eq.~(\ref{eq:S-opsN}).

We now assume that the perturbation preserves
a symmetry which forbids mixing between $J^x$ and $J^y$,
namely $\langle J^x J^y \rangle=0$.
Then the correlation function of the total spin can be
decoupled to $J^x$ and $J^y$ part.
\begin{eqnarray}
\langle S^+(t) S^-(0) \rangle  &=&
\frac{1}{8 \pi^2}
      \int dx_1
      \int dx_2
       \langle J^x_R(t,x_1) e^{i H (x_1 - x_2) } J^x_R(0,x_2) \rangle
  +    \langle J^x_L(t,x_1) e^{-i H (x_1 - x_2) } J^x_L(0,x_2) \rangle
\nonumber \\
&&  +     \langle J^x_R(t,x_1) e^{i H (x_1 + x_2) } J^x_L(0,x_2) \rangle
  +    \langle  J^x_L(t,x_1) e^{-i H (x_1 + x_2)} J^x_L(0,x_2) \rangle
\nonumber \\
&&
  + ( J^x \rightarrow J^y )
\label{eq:SpmJJ}
\end{eqnarray}
Since our effective field theory is $SU(2)$-symmetric, we may
freely rotate the $xyz$-axes. Thus, instead of calculating
correlation functions of $J^x$ we can calculate those of $J^z$,
{\em with perturbations also rotated correspondingly}. The same
applies to calculation of $J^y$ correlations.

The motivation for us to rotate the $xyz$-axes is that, $J^z$ is
expressed as a derivative of the boson field $\phi$ as in
eq.~(\ref{eq:JzRphi}). Thus the problem is reduced to the calculation
of the bosonic correlation function $\langle \phi \phi \rangle$.
The structure of the bosonic correlation function is well
established by the standard diagrammatic perturbation theory, and
the ESR lineshape is related to the boson self-energy as we will
show below. On the other hand, when the perturbation allows mixing
of $J^x$ and $J^y$ (in the original representation), there seems
no way to reduce the problem to the $\langle \phi \phi \rangle$
correlation function. In such cases, we do not know at present how
to construct the theory of ESR based on self-energy. Thus, below
we restrict ourselves to the situation in which $J^x$ and $J^y$ do
not mix, in the discussion of the self-energy approach. We remark
that there is no apparent difficulty in the application of the MK
formula even in  cases where the perturbation allows mixing of
$J^x$ and $J^y$.

As mentioned in Section~\ref{sec:perturbations}, we restrict
ourselves to the case where the perturbation does not contain an
oscillating factor $e^{i Hx}$. Then the contribution from the
cross terms such as $\langle J_R J_L \rangle$ vanish in
eq.~(\ref{eq:SpmJJ}), due to
 momentum conservation.
The correlation function thus reduces, upon Fourier transformation
to
\begin{equation}
\langle S^+ S^- \rangle (\omega) =
 \frac{1}{8 \pi^2}
 \left[
    \langle J^x_R J^x_R \rangle (\omega, - H)
    + \langle J^x_L J^x_L \rangle (\omega,  H)
    + \langle J^y_R J^y_R \rangle (\omega, -H)
    + \langle J^y_L J^y_L \rangle (\omega,  H)
\right]
\end{equation}
where $\langle J J \rangle (\omega, q)$ denotes the
correlation function at frequency $\omega$ and momentum $q$.
As we have discussed above, we now rotate the axes and
calculate $J^z$ correlation function instead of $J^x$ and $J^y$, to
obtain
\begin{eqnarray}
\lefteqn{\langle S^+ S^- \rangle (\omega) =} \nonumber \\
&& \frac{1}{8 \pi^2} \bigl[
    \langle J^z_R J^z_R \rangle_{x \rightarrow z} (\omega, H)
    + \langle J^z_L J^z_L \rangle_{x \rightarrow z} (\omega, - H)
    + \langle J^z_R J^z_R \rangle_{y \rightarrow z} (\omega, H)
    + \langle J^z_L J^z_L \rangle_{y \rightarrow z} (\omega, - H)
    \bigr],
\end{eqnarray}
where $\langle \rangle_{x\rightarrow z}$ means the correlation function
with the perturbation rotated $x \rightarrow z$.
Using eq.~(\ref{eq:JzRphi}) and~(\ref{eq:JzLphi}),
those correlation functions can
be written in terms of bosonic correlation function:
\begin{equation}
\langle S^+ S^- \rangle (\omega) =
 \frac{(\omega + H)^2}{4 \pi}
    \langle \phi \phi \rangle_{x \rightarrow z} (\omega, H)
+ \frac{(\omega + H)^2}{4 \pi}
    \langle \phi \phi \rangle_{y \rightarrow z}(\omega, H),
\label{eq:SSphi}
\end{equation}
where we have used the symmetry
$\langle \phi \phi \rangle (\omega, -H) =
\langle \phi \phi \rangle (\omega, H)$.
The above formula is useful if the perturbation (after the rotation)
is given by a Lagrangian density local in the boson field $\phi$.
If, for example, the Lagrangian density is local in terms
of the dual field $\tilde{\phi}$ after the rotation $y \rightarrow z$,
the second term in eq.~(\ref{eq:SSphi}) should be replaced by
\begin{equation}
\frac{(\omega + H)^2}{4 \pi}
\langle \tilde{\phi} \tilde{\phi} \rangle_{y \rightarrow z}(\omega, H).
\end{equation}

In fact, there is a subtlety in defining the current.
In the free boson theory without interactions, we have
\begin{eqnarray}
 \frac{\partial \phi}{\partial x} &=& \frac{\partial \tilde{\phi}}{\partial t},
\label{eq:phiphitilde}
\\
 \frac{\partial \phi}{\partial t} &=&
 -\frac{\partial \tilde{\phi}}{\partial x},
\label{eq:phiphitilde2}
\end{eqnarray}
so that we may represent the current operator in terms of either $\phi$
or $\tilde{\phi}$.
However, in the presence of the interaction,
we cannot define the dual fields $\phi$ and $\tilde{\phi}$
that satisfy both identities.
For example, let us take the Lagrangian density
\begin{equation}
{\cal L} = \frac{1}{2}(\partial_{\mu} \phi)^2 - \lambda \cos{\beta \phi},
\end{equation}
and {\em define} the dual field $\tilde{\phi}$ by
eq.~(\ref{eq:phiphitilde}).
Then, from the equation of motion, we find
\begin{equation}
 \partial_x \tilde{\phi}(t,x) + \partial_t \phi(t,x) =
 - \beta \lambda \int^t_{-\infty} \cos{\beta \phi}(t',x) dt' ,
\label{eq:phidiff}
\end{equation}
violating eq.~(\ref{eq:phiphitilde2}).

Thus it is not completely clear whether
the current operator should be written as
a derivative of $\phi$ or $\tilde{\phi}$.
However, upon Fourier transform, the ``difference term'' (right-hand side
of eq.~(\ref{eq:phidiff})) does not give a sharp peak.
(Recall that only the operators of conformal weight $(1,0)$ or $(0,1)$
produce a delta-function spectrum. Other operators give broad spectrum
given by eq.~(\ref{eq:Schulz}), even in the zeroth order.)
Moreover, the contribution from the difference term is suppressed
by a factor $\lambda^2$.
Therefore, the difference term would lead, at most,
only to a small and broad background.
In discussing the lineshape of the main resonance,
we can ignore the difference term and
focus on the derivative of either boson field $\phi$
or $\tilde{\phi}$.
For calculational convenience, we choose to use $\phi$ (or $\tilde{\phi}$)
if the interaction is given in terms of $\phi$ ($\tilde{\phi}$.)

Thus the problem of finding the ESR absorption spectrum is reduced to
the calculation of the correlation function of the boson field $\phi$.
We now make the Wick rotation and consider the corresponding Matsubara
Green's function defined by
\begin{equation}
\calG_{AB}(\tau) = - \frac{1}{Z}
 {\rm Tr} T_{\tau} \left[ A(\tau) B(0) \right] ,
\end{equation}
where $T_{\tau}$ is the ordering operator with respect to the
imaginary time $\tau$, and $A(\tau) \equiv e^{\tau \calH} A
e^{-\tau \calH}$. The standard diagrammatic perturbation theory
can be applied to the Matsubara Green's function. After obtaining
the Matsubara Green's function, we can analytically continue back
to  real time to obtain the retarded Green's function.

Provided that the Lagrangian is local in terms of the boson field,
its correlation function can be written in a self-energy form:
\begin{equation}
\calG_{\phi \phi} (\omega_n, q)
= \frac{-1}{ \omega_n^2 + q^2 + \Pi(\omega_n, q)},
\label{eq:GMboson}
\end{equation}
where $\calG$ is the (full) Matsubara Green's function,
$\omega_n$ is the Matsubara frequency,
and $\Pi(\omega_n,q)$ is the self-energy, namely
the sum of all one-particle irreducible diagrams.
Thus we obtain
\begin{equation}
\calG_{S^+ S^-}(\omega_n, q) \sim
 \frac{(i \omega_n + H)^2}{4 \pi}
    \frac{-1}{\omega_n^2 + H^2 + \Pi_x(\omega_n,H)}
+ \frac{(i \omega_n + H)^2}{4 \pi}
    \frac{-1}{\omega_n^2 + H^2 + \Pi_y(\omega_n,H)},
\end{equation}
where $\Pi_x$ and $\Pi_y$ are the self-energy in the Matsubara
formalism, respectively for
$\langle \phi \phi \rangle_{x \rightarrow z}$ and
$\langle \phi \phi \rangle_{y \rightarrow z}$.
This gives, upon the analytic continuation, the retarded Green's function
\begin{equation}
\calG^R_{S^+ S^-}(\omega, q) \sim
 \frac{(\omega + H)^2}{4 \pi}
    \frac{1}{\omega^2 - H^2 - \Pi^R_x(\omega,H)}
+ \frac{(\omega + H)^2}{4 \pi}
    \frac{1}{\omega^2 - H^2 - \Pi^R_y(\omega,H)},
\label{eq:ESRandPi}
\end{equation}
where the ``self-energy'' $\Pi^R_{\alpha}$ ($\alpha=x,y$)
is defined by the analytic continuation
\begin{equation}
 \Pi^R_{\alpha}(i \omega_n,q) = \Pi_{\alpha}(\omega_n,q)
\end{equation}
for $\omega_n >0$.

First let us check what we obtain in the absence of the
perturbation. Then $\Pi^R_x = \Pi^R_y =0$ so that the Green's function
has a pole at $\omega = H$:
\begin{equation}
G^R_{S^+S^-}(\omega) \sim \frac{H}{\pi} \frac{1}{\omega - H + i 0}.
\label{eq:SSfree}
\end{equation}
This means that we have a completely sharp resonance at the Zeeman
energy $\omega = H$ as expected, in agreement with the equation of
motion. The residue $H/\pi$ at the pole of the Green's function
gives the intensity of the resonance. This is also consistent with
the exact result from the original spin chain.
\begin{equation}
G^R_{S^+S^-}(\omega) =
    -i \int_0^{\infty} dt \langle [ S^+(t) , S^-(0) ] \rangle
= \frac{2m}{\omega - H + i 0} ,
\end{equation}
where $m$ is the magnetization.
For small field $H$, the magnetization is given by $m = \chi_u H$,
where the uniform susceptibility is
\begin{equation}
\chi_u = \frac{1}{2 \pi}
\label{eq:chiu}
\end{equation}
in the low-temperature limit, ignoring the effect of the isotropic
marginal operator\cite{Eggert}.
(We remind the reader that we have been setting $v=1$.)
Thus we obtain the amplitude $2m = H/\pi$, in agreement with
eq.~(\ref{eq:SSfree}).

A symmetry breaking perturbation $\calH'$ would give non-vanishing
boson self-energy $\Pi_x,\Pi_y$. This changes the ESR lineshape.
Near the resonance $\omega \sim H$, we can write
\begin{equation}
\calG^R_{S^+ S^-} (\omega) =
 \frac{H}{2 \pi} \frac{1}{\omega - H - \frac{1}{2H} \Pi^R_x(\omega,H)}
+ \frac{H}{2 \pi} \frac{1}{\omega - H - \frac{1}{2H} \Pi^R_y(\omega,H)}
\end{equation}
If the self-energy changes smoothly around the resonance $\omega
\sim H$, we may regard the self-energy as being constant in a
frequency range sufficiently close to the center of resonance.
Then, within this range, the lineshape is given by a Lorentzian,
and the real and imaginary parts of the self-energy give the shift
and width of the ESR, respectively. The linewidth is given by
\begin{equation}
    \eta = \frac{-1}{2H} {\rm Im} {\Pi^R_{\alpha}(H,H)},
\label{eq:selfwidth}
\end{equation}
while the shift is
\begin{equation}
    \Delta \omega = \frac{1}{2H} {\rm Re}{\Pi^R_{\alpha}(H,H)},
\label{eq:selfshift}
\end{equation}
for $\alpha = x,y$.
In general, the signal could be superposition of two Lorentzian
spectra corresponding to $\Pi^R_x$ and $\Pi^R_y$.
However, in the concrete cases we study in the present
paper, $\Pi^R_x$ and $\Pi^R_y$ are equal; thus
a single Lorentzian lineshape is predicted.

Therefore we have successfully formulated the theory
of ESR without any particular assumption on the lineshape.
The self-energy is usually a smooth function of $\omega$
near $\omega \sim H$ for finite $H$ except for the smooth 
weak background discussed below Eq. (\ref{eq:phidiff}); we have given
a microscopic foundation for the Lorentzian lineshape
which is assumed a priori in the MK approach.
Application of the present self-energy formalism to
two cases relevant to experiments
will be discussed in the following sections.
However, precisely speaking,
our approach
is only formulated ignoring the isotropic marginal
operator, which is generally present in the effective theory
of the Heisenberg antiferromagnetic chains.
Some discussions on the effects of the isotropic marginal
operator will be given in Section~\ref{sec:logcorr}.

Comparing with the assumption~(\ref{eq:lor}) used in
our derivation of the
MK formula in Appendix~\ref{sec:deriveMK}, it is obvious that
the MK formula and the self-energy approach are closely related.
Namely, $\Sigma$ introduced in eq.~(\ref{eq:lor}) corresponds to
$\Pi^R/(2H)$ if they vary smoothly around the resonance. The
important difference is that it is an assumption that the Green's
function can be written as in eq.~(\ref{eq:lor}) with a smooth
$\Sigma$ whereas we can prove eq.~(\ref{eq:GMboson}) using
the diagrammatic perturbation theory. The self-energy, $\Pi$, is given
by the sum of all one-particle irreducible Feynman diagrams as in
proven in any book on field theory. In this way, our self-energy
formulation effectively gives a proof of the Lorentzian
form~(\ref{eq:lor}) which is often assumed without a microscopic
foundation. We emphasize that, although eq.~(\ref{eq:lor}) may
appear innocent, it is a rather strong assumption and is far from
trivial.
When the lineshape turns out to be Lorentzian, the results
must agree between the MK and self-energy approaches,
if the correlation functions are evaluated correctly.
This will be verified for a few cases in Sections~\ref{sec:anisMK},
\ref{sec:stagMK} and~\ref{sec:stagshift}. 
On the other hand, while the validity of the MK formula
is limited to the lowest order perturbation theory,
the self-energy formulation allows us to
go beyond that.
In fact, we will make a non-perturbative analysis
of the lineshape, based on the self-energy formalism,
in Section~\ref{sec:vltemp}.

We note that assumptions similar to eq.~(\ref{eq:lor}) have been
made in the literature for different problems.
Sometimes the (counterpart of) assumed $\Sigma$ in eq.~(\ref{eq:lor})
is referred to as the memory function.
For example, Giamarchi\cite{Giamarchi} studied
the conductivity of the TL liquid with the bosonization method.
His discussion was rather closely related to our analysis of ESR
in the present paper.
(See also Ref.~\onlinecite{Goetze}.)
In fact, he calculated the ac conductivity of a TL liquid
by evaluating the memory function with the field theory.
This is quite similar to a field-theory calculation of the MK formula
for ESR, which we will discuss in later sections.

We could also apply our self-energy approach to problems such as
the conductivity of TL liquids.
This might be useful
for providing a more rigorous justification
and a possibility to go beyond the lowest order perturbation theory.
The possible breakdown of the MK 
formula, in the context of the conductivity of a TL liquid, 
was discussed by Giamarchi and Millis\cite{Giamarchi2}.

\section{Exchange anisotropy perpendicular to the magnetic field}
\label{sec:anis-x}

Now we consider the exchange anisotropy with the axis
perpendicular to the applied magnetic field.
Let us take the axis of the anisotropy as the $x$-axis.
In the low-energy effective theory, {\em at zero uniform field},
the anisotropy term is given as
\begin{eqnarray}
{\cal L}_a &=& - \lambda J^x_R J^x_L  \nonumber \\
&=& - \frac{\lambda}{2}(J^x_R J^x_L - J^y_R J^y_L )
 + \frac{\lambda}{2}J^z_R J^z_L
 + \frac{\lambda}{2}\vec{J}_R  \cdot \vec{J}_L .
\label{eq:anis-eff}
\end{eqnarray}
Here the parameter $\lambda$, which is proportional to $\delta$
for a small $\delta$, is the same as the one introduced
in eq.~(\ref{eq:anisz-eff}).
The last term $\vec{J}_R \cdot \vec{J}_L$ of the second line is the
isotropic marginal operator, which does not affect the resonance
directly and will thus be ignored in the following.

Now let us include the effects~(\ref{eq:phishift})
of the applied uniform field $H$.
The first and second terms in eq.~(\ref{eq:anis-eff}) are transformed into
\begin{equation}
{\cal L}_a =
 - \frac{\lambda}{2}(J^x_R J^x_L - J^y_R J^y_L )
 + \frac{\lambda}{2}J^z_R J^z_L
 + \frac{\lambda H}{2 \sqrt{2}} (J^z_R + J^z_L)
 + \frac{\lambda H^2}{4} .
\label{eq:anis-eff2}\end{equation} Fortunately, there is no
oscillating factor $e^{iHx}$ here. The last constant term has no
effect in the following, and will be ignored. The third term
represents the additional magnetic field of $+ \pi \lambda H$.
(Compare with eq.~(\ref{eq:aniszLag}).) This is again absorbed by
a renormalization of the uniform field $H$, giving the shift of
$\pi \lambda H$.

The remaining problem then is to study the effect of the
perturbation
\begin{equation}
{\cal L}'_a =
 - \frac{\lambda}{2}(J^x_R J^x_L - J^y_R J^y_L - J^z_R J^z_L )
\end{equation}
The first two terms corresponds to an interaction in terms of the
boson field $\tilde{\phi}$, and the problem cannot be reduced to a
free field theory. Thus it is not possible to calculate the ESR
absorption spectrum directly as we have done for the exchange
anisotropy parallel to the magnetic field in
Section~\ref{sec:anis-z}. Therefore, we will employ the
self-energy approach developed in Section~\ref{sec:self-energy}.

\subsection{Self-energy approach}

Because the anisotropy considered here breaks the rotational symmetry
in the $xy$-plane, we expect a polarization dependence.
Thus let us consider the correlation function of $S^x$ and $S^y$
separately.
Under the magnetic field, $S^{x,y}$ at zero momentum are
expressed as
\begin{eqnarray}
S^x &=&
\frac{J^+_R(H) + J^+_L(-H) + J^-_R(-H) + J^-_L(H)}{2 \sqrt{8 \pi^2}}, \\
S^y &=&
\frac{J^+_R(H) + J^+_L(-H) - J^-_R(-H) - J^-_L(H)}{2 i \sqrt{8 \pi^2}} .
\end{eqnarray}
We emphasize here that, under the magnetic field, $S^x$ is related
to both current operators $J^x$ and $J^y$.
The original spin operator and the current operator are quite different
objects.

Absorbing the third term in~(\ref{eq:anis-eff2}) as a
renormalization of the magnetic field, the perturbation respects
the symmetry $J^x \rightarrow J^x, J^y \rightarrow - J^y, J^z
\rightarrow - J^z$. Thus the cross term $\langle J^x J^y \rangle$
vanishes in this case, allowing us to proceed with the rotation
trick described in Section~\ref{sec:self-energy}. Namely,
\begin{eqnarray}
\langle S^x S^x \rangle &=&
\frac{1}{32\pi^2}
\bigl[
\langle J^x_R J^x_R \rangle + \langle J^x_R J^x_L \rangle
+ \langle J^x_L J^x_R \rangle + \langle J^x_L J^x_L \rangle
\nonumber \\
&& + \langle J^y_R J^y_R \rangle + \langle J^y_R J^y_L \rangle +
\langle J^y_L J^y_R \rangle + \langle J^y_L J^y_L \rangle  \bigr]
(q=H)
\nonumber \\
&& + \frac{1}{32\pi^2} \bigl[ \ldots \bigr] (q= -H)
\nonumber \\
&=&
\frac{1}{32\pi^2}
\bigl[
\langle J^z_R J^z_R \rangle_x + \langle J^z_R J^z_L \rangle_x
+ \langle J^z_L J^z_R \rangle_x + \langle J^z_L J^z_L \rangle_x
\nonumber \\
&& + \langle J^z_R J^z_R \rangle_y + \langle J^z_R J^z_L \rangle_y
+ \langle J^z_L J^z_R \rangle_y + \langle J^z_L J^z_L \rangle_y
\bigr] (q=H)
\nonumber \\
&& + \frac{1}{32\pi^2} \bigl[ \ldots \bigr] (q= -H) ,
\label{eq:SxSx-x}
\\
\langle S^y S^y \rangle &=&
\bigl[
 \langle J^x_R J^x_R \rangle - \langle J^x_R J^x_L \rangle
- \langle J^x_L J^x_R \rangle + \langle J^x_L J^x_L \rangle
\nonumber \\
&& + \langle J^y_R J^y_R \rangle - \langle J^y_R J^y_L \rangle - 
\langle J^y_L J^y_R \rangle + \langle J^y_L J^y_L \rangle \bigr]
(q=H)
\nonumber \\
&& + \bigl[ \ldots \bigr] (q= -H)
\nonumber \\
&=& \bigl[ \langle J^z_R J^z_R \rangle_x - \langle J^z_R J^z_L
\rangle_x - \langle J^z_L J^z_R \rangle_x + \langle J^z_L J^z_L
\rangle_x
\nonumber \\
&&
+ \langle J^z_R J^z_R \rangle_y - \langle J^z_R J^z_L \rangle_y
- \langle J^z_L J^z_R \rangle_y + \langle J^z_L J^z_L \rangle_y \bigr] (q=H)
\nonumber \\
&&+ \bigl[ \ldots \bigr] (q= -H) .
\label{eq:SySy-x}
\\
\langle S^x S^y \rangle &=&
\frac{1}{32\pi^2 i}
\bigl[
 \langle J^x_R J^x_R \rangle - \langle J^x_R J^x_L \rangle
+ \langle J^x_L J^x_R \rangle - \langle J^x_L J^x_L \rangle
\nonumber \\
&& + \langle J^y_R J^y_R \rangle + \langle J^y_R J^y_L \rangle
- \langle J^y_L J^y_R \rangle - \langle J^y_L J^y_L \rangle \bigr]
(q=H)
\nonumber \\
&& - \bigl[ \ldots \bigr] (q= -H)
\nonumber \\
&=& \bigl[ \langle J^z_R J^z_R \rangle_x - \langle J^z_R J^z_L \rangle_x
+ \langle J^z_L J^z_R \rangle_x - \langle J^z_L J^z_L \rangle_x
\nonumber \\
&&
+ \langle J^z_R J^z_R \rangle_y + \langle J^z_R J^z_L \rangle_y
- \langle J^z_L J^z_R \rangle_y - \langle J^z_L J^z_L \rangle_y \bigr] (q=H)
\nonumber \\
&& - \bigl[ \ldots \bigr] (q= -H) .
\end{eqnarray}
Here $\langle \rangle_x$ and $\langle \rangle_y$
means the expectation value in
the presence of the (rotated) perturbation
$\lambda/2 [ J^z_L J^z_R - (J^x_L J^x_R + J^y_L J^y_R)]$ and
$\lambda/2 [ (J^x_L J^x_R - J^y_L J^y_R) - J^z_L J^z_R]$,
respectively.
Fortunately, these can be written
in terms of either $\phi$ or $\tilde{\phi}$:
\begin{eqnarray}
J^z_L J^z_R - (J^x_L J^x_R + J^y_L J^y_R) &=&
- \pi (\partial_{\mu} \phi)^2 - 2 \cos{\sqrt{8\pi}\phi},
\\
(J^x_L J^x_R - J^y_L J^y_R) - J^z_L J^z_R &=&
\pi (\partial_{\mu} \tilde{\phi})^2 + 2 \cos{\sqrt{8\pi} \tilde{\phi}}.
\end{eqnarray}
The $(\partial_{\mu} \phi)^2$ term gives a renormalization of
the radius $R$.
However, in the lowest order of the perturbation
theory, its effect is negligible on the boson correlation function
$\langle \phi \phi \rangle$ and thus will be dropped in the
following.

Thus, in evaluating $\langle J^z J^z \rangle_x$ we will
represent the current operator
$J^z$ as a derivative of $\phi$,
so that the problem is reduced to the correlation function
of the fundamental boson field $\phi$ in the presence of
the interaction in terms of $\phi$.
On the other hand, in evaluating $\langle J^z J^z \rangle_y$,
we will express the current $J^z$ by of $\tilde{\phi}$
for $\langle J^z J^z \rangle_y$.

As a result, we have
\begin{eqnarray}
{\calG^R_{J^z_R J^z_R}}^x (\omega,q)&=&
 \pi  \frac{(\omega - q)^2}{\omega^2 - q^2 - \Pi^R(\omega,q) } ,
\\
{\calG^R_{J^z_R J^z_L}}^x (\omega,q)&=&
 \pi \frac{\omega^2 - q^2}{\omega^2 - q^2 - \Pi^R (\omega,q)} ,
\\
{\calG^R_{J^z_L J^z_L}}^x (\omega,q)&=&
\pi \frac{(\omega +  q)^2}{\omega^2 - q^2 - \Pi^R (\omega,q)},
\\
{\calG^R_{J^z_R J^z_R}}^y (\omega,q)&=&
\pi  \frac{(\omega - q)^2}{\omega^2 - q^2 - \Pi^R(\omega,q) },
\\
{\calG^R_{J^z_R J^z_L}}^y (\omega,q)&=&
\pi \frac{\omega^2 - q^2}{\omega^2 - q^2 - \Pi^R (\omega,q)},
\\
{\calG^R_{J^z_L J^z_L}}^y (\omega,q)&=& \pi \frac{(\omega +
q)^2}{\omega^2 - q^2 - \Pi^R (\omega,q)},
\label{eq:Gfinal}
\end{eqnarray}
where ${\calG^R}^{\alpha}$ ($\alpha=x,y$) is the
retarded Green's function defined by the expectation value
$\langle \ldots \rangle_{\alpha}$, $\Pi^R(\omega,q)$ is the
self-energy for the boson field $\phi$ in the presence of the
interaction $- \lambda \cos{\sqrt{8\pi}\phi}$ (or the self-energy
for the boson field $\tilde{\phi}$ in the presence of $\lambda
\cos{\sqrt{8\pi} \tilde{\phi}}$, but this is identical.) Plugging
these into eqs.~(\ref{eq:SxSx-x}),(\ref{eq:SySy-x}), we obtain
\begin{eqnarray}
\calG^R_{xx} (\omega) &= &
\frac{H^2}{2 \pi} \frac{1}{\omega^2 - H^2 -  \Pi^R(\omega,H)} ,
\label{eq:GRxxex}
\\
\calG^R_{yy} (\omega) &= &
\frac{\omega^2}{2 \pi} \frac{1}{\omega^2 - H^2 - \Pi^R(\omega,H)},
\label{eq:GRyyex}
\\
\calG^R_{xy} = - \calG^R_{yx}(\omega) &= &
i \frac{\omega H}{2 \pi} \frac{1}{\omega^2 - H^2 - \Pi^R(\omega,H)},
\label{eq:GRxyex}
\end{eqnarray}
where $\calG^R_{\alpha \beta}$ is
the retarded Green's functions of the spin operators $S^{\alpha}$
and $S^{\beta}$, as defined in eq.~(\ref{eq:defgr}).

For a direction $\alpha$ in the $xy$-plane,
\begin{equation}
\calG^R_{\alpha \alpha} =
\frac{H^2 \cos^2{\Phi} + \omega^2 \sin^2{\Phi} }{\pi}
\frac{1}{\omega^2 - H^2 -  \Pi^R(\omega,H)},
\end{equation}
where $\Phi$ is the angle between $x$ and $\alpha$ directions,
namely the angle between the anisotropy axis and the polarization of the
electromagnetic wave.

As a result, for any directions of the polarization perpendicular to
the magnetic field, the ESR lineshape is Lorentzian with
the width $- {\rm Im}\Pi^R(H,H)/(2H)$.
However, the lineshape has some angle dependence through
the numerator $H^2 \cos^2{\theta} + \omega^2 \sin^2{\theta}$.
In fact, the present result is consistent with the exact and
rigorous relation~(\ref{eq:pol-dependence}) for
original spin model.
This serves as a consistency check of our field-theory approach.

\begin{figure}
\begin{center}
\epsfysize=10cm
\epsfbox{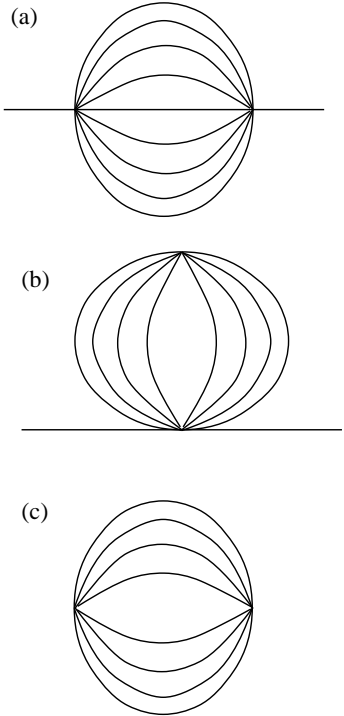}
\caption{
Three types of Feynman diagrams appearing in the perturbative expansion.
(a),(b) and (c) correspond to
the first, second and third terms in eq.~(\protect\ref{eq:GFexpand}),
respectively.
The disconnected diagram (c) is canceled by the correction to the
partition function; the ``tadpole'' diagram (b) does not contribute
to the imaginary part of the self-energy  (i.e. the linewidth)}
\label{fig:diagram}
\end{center}
\end{figure}

Now let us calculate the self-energy $\Pi$ of boson field $\phi$
in the presence of interaction $\lambda \cos{\sqrt{8 \pi}\phi}$.
It is easy to see the first order perturbation to the boson
correlation function vanishes due to symmetry. The second order
perturbation to the boson correlation function does not vanish and
can be calculated by the diagrammatic expansion (ie. Wick's
theorem). The second-order term in the boson correlation function
is related to
\begin{eqnarray}
\frac{\lambda^2}{2 \cdot 4}
\langle \phi(1)
e^{i \sqrt{8 \pi} \phi(2)} e^{- i \sqrt{8 \pi} \phi(3)}
\phi(4) \rangle
&=&
\frac{\lambda^2}{8}
\sum_{n,m} \frac{(i \sqrt{8\pi})^n}{n!} \frac{(-i \sqrt{8\pi})^m}{m!}
    \langle \phi(1) : \phi^n(2) : :\phi^m(3): \phi(4) \rangle
\\
&=&
\pi \lambda^2 \langle \phi(1) \phi(2) \rangle
\langle e^{i \sqrt{8 \pi} \phi(2)} e^{- i \sqrt{8 \pi} \phi(3)} \rangle
\langle \phi(3) \phi(4) \rangle
+ ( 2 \leftrightarrow 3 ) \nonumber \\
&&
- \pi  \lambda^2 \langle \phi(1) \phi(2) \rangle
\langle e^{i \sqrt{8 \pi} \phi(2)} e^{- i \sqrt{8 \pi} \phi(3)} \rangle
\langle \phi(2) \phi(4) \rangle
+ ( 2 \leftrightarrow 3 ) \nonumber \\
&& + \pi \lambda^2
\langle \phi(1) \phi(4) \rangle
\langle e^{i \sqrt{8 \pi} \phi(2)} e^{- i \sqrt{8 \pi} \phi(3)} \rangle
\label{eq:GFexpand}
\end{eqnarray}
The three terms here represent contributions form different kinds
of Feynman diagrams, as shown in Fig.~\ref{fig:diagram}.
The second type of the term
$ -  \langle \phi(1) \phi(2) \rangle
\langle e^{i \sqrt{8 \pi} \phi(2)} e^{- i \sqrt{8 \pi} \phi(3)} \rangle
\langle \phi(2) \phi(4) \rangle
+ ( 2 \leftrightarrow 3 )$
represents the ``tadpole'' type Feynman diagram
(Fig.~\ref{fig:diagram} (b)),
while
the last term corresponds to a disconnected Feynman diagram
(Fig.~\ref{fig:diagram} (c)),
which is canceled by the correction to the partition function.

In fact, there is a similar contribution from
$e^{-i \sqrt{8 \pi} \phi(2)} e^{+ i \sqrt{8 \pi} \phi(3)}$
besides the above, and one has to integrate the coordinates $2$
and $3$ over Euclidean space-time.
As a result, we obtain the self-energy in the lowest order
($O(\lambda^2)$) of the perturbation as
\begin{equation}
\Pi(\omega_n,q) =
4 \pi \lambda^2 [ G_{(1,1)}(\omega_n,q) - G_{(1,1)}(0,0) ],
\end{equation}
where $G_{(1,1)}$ is the Matsubara Green's function of the
operator of the conformal weight $(1,1)$ in the free boson theory.
These two terms come from type (a) and (b) Feynman diagrams
in Fig.~\ref{fig:diagram}, respectively.
Analytic continuation back to  real time leads to
\begin{equation}
\Pi^R(\omega,q) =
4 \pi \lambda^2 [ G^R_{(1,1)}(\omega,q) - G^R_{(1,1)}(0,0) ],
\end{equation}
where $G^R_{(1,1)}$ is the retarded Green's function
corresponding to the Matsubara Green's function $G_{(1,1)}$.
Its imaginary part
can be derived by taking the limit $\Delta,\bar{\Delta} \rightarrow 1$
in eq.~(\ref{eq:Schulz}):
\begin{equation}
{\rm Im}[ - G^R_{(1,1)}(\omega,q) ] = \frac{\pi^2}{8} (\omega^2 - q^2)
 \left[ \coth{\frac{\omega+q}{4T}} + \coth{\frac{\omega-q}{4T}}\right] .
\label{eq:GR11}
\end{equation}
The imaginary part then reads
\begin{equation}
- {\rm Im} \Pi^R (H,H) = 4 \pi^3 \lambda^2 HT,
\end{equation}
giving the width
\begin{equation}
\eta = 2 \pi^3 \lambda^2 T
\label{eq:xxwidth}
\end{equation}
Again, this is consistent with the scaling analysis~(\ref{eq:deltascale}).
The real part is proportional to $(\omega^2 - q^2)$, which corresponds
to a wavefunction renormalization, and does not lead to any
shift at $O(\lambda^2)$.
In any case, there is a shift of $O(\lambda)$ discussed above,
which is dominant.

To summarize, the exchange anisotropy with the axis
perpendicular to the applied field gives the following effects
on paramagnetic ESR.
\begin{description}
\item[shift] $ + \pi \lambda H \propto H \delta $
\item[width] $2 \pi^3 \lambda^2 T \propto (\delta/J)^2 T$
\end{description}
Comparing to the result for the exchange anisotropy with the axis
parallel to the applied field, the width obtained here is half of
the result~(\ref{eq:zzwidth}) for the parallel case. This can be
understood naturally with the MK formula as we will discuss in the
next subsection. On the other hand, the shift takes opposite sign
and the absolute value is half of that in the parallel case.

\subsection{MK approach}
\label{sec:anisMK}

The lineshape is shown to be Lorentzian
in the two cases discussed above (exchange anisotropy parallel and
perpendicular to the applied field),
up to a possible broad background of $O(\lambda^2)$.
Thus the MK formula is expected to be also valid for these cases.
In order to check
consistency of our field-theory approach, here we study the same
problem with the MK formula.

Let us consider the exchange anisotropy parallel to the applied
field considered in Section~\ref{sec:anis-z}. We may apply the
MK formula to the spin chain Hamiltonian and then take the
continuum limit, but taking the continuum limit first and then
apply the MK formula turns out to be simpler. Absorbing the second
term of the effective perturbation~(\ref{eq:aniszLag}) into a
renormalization of the magnetic field, we need to consider the
effect of the perturbation $\calH' \sim \lambda \int dx J^z_L
J^z_R$.

First we have to obtain the commutator~(\ref{eq:defA}) appearing in
the MK formula.
The total spin raising/lowering operator $S^{\pm}$ in the continuum
limit is given from eq.~(\ref{eq:S-opsN}) as
\begin{equation}
    S^{\pm} =
        \frac{1}{\sqrt{8 \pi^2}}
        \int dx ( J^{\pm}_R e^{\pm i Hx} + J^{\pm}_L e^{\mp i Hx} ).
\label{eq:totalSpmJ}
\end{equation}
Using the standard commutation relation among the currents,
the commutator $\calA$ is given by
\begin{equation}
\calA = i [ \calH' , S^+ ] =
 i \lambda \int dx
        [J_L^z(x) J_R^+(x) e^{iHx} + J_R^z(x) J_L^+(x) e^{-iHx}].
\label{eq:Apar} \end{equation} $J_L^z J_R^+$ and $J_R^z J_L^+$ are
primary fields with the conformal weight $(1,1)$. Thus, from the
MK formula~(\ref{eq:MKwidth}) we obtain the linewidth
\begin{equation}
 \eta = \frac{2 \lambda^2}{\chi_u H} {\rm Im}[-G^R_{(1,1)}(H,H)]
\end{equation}
The Green's function is what
we have already considered in~(\ref{eq:GR11}), and thus we
obtain the width
\begin{equation}
 \eta = \frac{2 \lambda^2}{\chi_u} \pi^2 T .
\end{equation}
Using eq.~(\ref{eq:chiu}) again (recall we have set $v=1$),
\begin{equation}
 \eta = 4 \pi^3 \lambda^2 T .
\end{equation}
This indeed agrees exactly with the result~(\ref{eq:zzwidth})
obtained by quite a different approach.  We remark that a similar 
derivation of a similar formula for the ac conductivity of a 
TL liquid was given earlier by Giamarchi\cite{Giamarchi}.

Next let us consider the exchange anisotropy perpendicular to the
applied field. Absorbing the third term in~(\ref{eq:anis-eff2})
into  the renormalization of the magnetic field, the perturbation
to be considered is $\calH' = (\lambda/2)\int dx (J^x_L J^x_R -
J^y_L J^y_R - J^z_L J^z_R) $. Consequently, the commutator becomes
\begin{equation}
\calA = i [ \calH' , S^+ ] =
 i \frac{\lambda}{2} \int dx
 [(J_L^-(x) J_R^z(x) + J^z_L J^+_R) e^{iHx} +
 (J_L^z(x) J_R^-(x) + J^+_L J^z_R) e^{-iHx}] ,
\label{eq:Aperp}\end{equation} This leads to
\begin{equation}
 \eta = \frac{\lambda^2}{\chi_u H} {\rm Im}[-G^R_{(1,1)}(H,H)]
 =  2 \pi^3 \lambda^2 T ,
\end{equation}
where we have used the susceptibility~(\ref{eq:chiu})  in the
second equality. Again we have found an exact agreement with the
self-energy approach~(\ref{eq:xxwidth}). The ratio $2$ of the
width between the parallel case~(\ref{eq:zzwidth}) and the
perpendicular case~(\ref{eq:xxwidth}) is simply understood in this
approach.  It arises from the factor of 1/2 and the presence of
twice as many terms in Eq. (\ref{eq:Aperp}) as compared to Eq.
(\ref{eq:Apar}). In fact, such an angle dependence also holds at
higher temperature and has been discussed in the literature, for
example in Refs.~\onlinecite{Date,Yamada:KCuF3}.


\subsection{Effect of the marginal isotropic operator: logarithmic correction}

\label{sec:logcorr}

The Hamiltonian of the Heisenberg antiferromagnetic chain with a
small anisotropy in the $z$ direction can be written as
\begin{equation}
    \calH = \calH_0 -
        \left[
            g^x (J^x_R J^x_L + J^y_R J^y_L )
            +g^z J^z_R J^z_L
        \right],
\end{equation}
where we ignored the applied field $H$,
which will be considered later.
Here we can rewrite the perturbation as
\begin{equation}
g^x \vec{J}_L \cdot \vec{J}_R + (g^z - g^x) J^z_L J^z_R ,
\end{equation}
where the first term is the isotropic marginal operator.
The second term gives the anisotropic interaction
$\lambda = - g^z + g^x$.

As is now well known, the isotropic marginal perturbation
exists in the low-energy effective theory of the Heisenberg
antiferromagnetic chain, giving several effects such as
the logarithmic correction to the magnetic susceptibility\cite{Eggert}
at low temperature.
While it has a simple form
$\vec{J}_L \cdot \vec{J}_R$ at $H=0$, it becomes complicated
if we include the effect of the applied field $H$.
It introduces complications such as the momentum non-conservation
in the effective theory and the mixing of $J^x$ and $J^y$, thereby
invalidating the simple self-energy approach discussed in
Section~\ref{sec:self-energy}.
Thus we actually have no microscopic derivation of the Lorentzian
lineshape in the presence of the isotropic marginal operator, at present.
On the other hand, the operator by itself, being isotropic, does not
directly affect the linewidth.
Since the isotropic marginal coupling constant renormalizes to zero,
 we may expect the Lorentzian lineshape
is basically unaffected by its presence.
It does, however, indirectly affect the linewidth through the
renormalization of the anisotropic perturbation as we discuss in
the following.

As discussed in Ref.~\onlinecite{Affleck:log}, the coupling
constants $g^x$ and $g^z$ are renormalized by the Kosterlitz-Thouless
type RG flow.
The solution of the RG equation (for $H=0$) in the lowest order gives
\begin{eqnarray}
g^x &=& \frac{\epsilon}{4 \pi} \frac{1}{\sinh{(\epsilon \ln{r}})} \\
g^z &=& \frac{\epsilon}{4 \pi} \coth{(\epsilon \ln{r})},
\end{eqnarray}
where $r$ is the scale variable ($\propto J/T$) and $\epsilon$ is a constant,
which determines the crossover scale. [This solution is valid only
if the infrared (IR) limit is a massless free boson theory, namely
if $\delta < 0$. We proceed by assuming this case; the final
result on the ESR linewidth should be valid also for $\delta>0$.]
In the IR limit $r \rightarrow \infty$, $g^x =0$ and
\begin{equation}
g^z(\infty) = \frac{\epsilon}{4 \pi}.
\end{equation}
This corresponds to a renormalized free boson Lagrangian $(1 - 2
\pi g^z (\infty) ) (\partial_{\mu} \phi)^2 / 2$, which leads to
the critical exponent $\eta_z = 1 - 2 \pi  g^z(\infty)$, where
$\langle S^z(r) S^z(0) \rangle \sim r^{- \eta_z}$.

On the other hand, the critical exponent in the low-energy limit
of the Heisenberg XXZ model has been obtained from the Bethe
Ansatz exact solution. For the Heisenberg model with an exchange
anisotropy
\begin{equation}
\calH = \sum_j J (S^x_j S^x_{j+1} + S^y_j S^y_{j+1}) +
    (J + \delta) S^z_j S^z_{j+1} ,
\label{eq:XXZ}
\end{equation}
it is known that
\begin{equation}
\eta_z = \frac{1}{2\pi R^2}=
 1 - \frac{1}{\pi} \cos^{-1}{[ 1 + \frac{\delta}{J} ]},
\label{eq:Rofdelta}
\end{equation}
for a negative $\delta$. Combining these results, we obtain, for
small $\epsilon$, $\delta$,
\begin{equation}
\epsilon = \frac{1}{\pi} \sqrt{ \frac{-8 \delta}{J}} .
\end{equation}
Since the isotropic part $ g^x \sum_{\alpha} J^{\alpha}_L
J^{\alpha}_R $ commutes with $S^+$, the important perturbation is
the ``asymmetric part'' $g^z - g^x$. In the intermediate scale $r
\ll e^{1/\epsilon}$, which would be relevant to  ESR for a weak
anisotropy,
\begin{equation}
    \lambda = - g^z + g^x = \frac{1}{8 \pi} \epsilon^2 \ln{r}
                  =   \frac{\ln{r}}{\pi^3} \frac{\delta}{J} .
\end{equation}
This corresponds to the coefficient $\lambda$ introduced in
eq.~(\ref{eq:anisz-eff}). The larger of the temperature $T$ or the
applied field $H$ imposes the cutoff of the RG flow, and thus the
scale factor $r$ should be replaced by $J/{\rm max}(T,H)$.

In the present discussion, the uniform field $H$ appears only
as a cutoff scale imposed on the RG flow at zero field.
Thus, to this order, the renormalization of the coupling constant
$\lambda$ applies to arbitrary direction of the anisotropy
relative to the applied field.
Therefore we conclude the low-temperature asymptotic behavior
of the linewidth and shift to be
\begin{eqnarray}
\eta &=&  \frac{4}{\pi^3} \left( \frac{\delta}{J} \right)^2
    \left( \ln{ \frac{J}{{\rm max}(T,H)}} \right)^2 T ,
\label{eq:widthdeltaz}
\\
\Delta \omega &=& -\frac{2}{\pi^2} \frac{\delta}{J}
                 \ln{ \frac{J}{{\rm max}(T,H)}},
\end{eqnarray}
if the anisotropy axis is parallel to the applied field.
They are
\begin{eqnarray}
\eta &=&  \frac{2}{\pi^3} \left( \frac{\delta}{J} \right)^2
    \left( \ln{ \frac{J}{{\rm max}(T,H)}} \right)^2 T ,
\label{eq:widthdeltax}
\\
\Delta \omega &=& \frac{1}{\pi^2} \frac{\delta}{J}
                 \ln{ \frac{J}{{\rm max}(T,H)}},
\end{eqnarray}
if the anisotropy axis is perpendicular to the applied field.
The shift depends on the sign of the anisotropy.
When comparing with experiments or existing literature,
it should be recalled that we discuss the shift in frequency
(for a fixed field $H$) while usually a shift in the resonance
field for a fixed frequency $\omega$ is studied.
For example, in the presence of the dipolar interaction, which
corresponds to {\em negative} $\delta$,
when the field is applied parallel to the chain (ie. anisotropy) axis
we obtain a positive frequency shift, namely a negative shift in the
resonance field. The shift is in the opposite direction when the
applied field is perpendicular to the chain axis.
These conclusions are qualitatively consistent with
the literature.~\cite{NT,Miya-ESR}

\subsection{Comparison with experiments}

In this paper, we have not calculated the ESR lineshape for a
general relative direction between the anisotropy axis and the
magnetic field, let alone more complicated anisotropy of general
form. However, the results~(\ref{eq:zzwidth}), (\ref{eq:xxwidth})
together with the scaling argument~(\ref{eq:deltascale}) imply
that the linewidth due to the exchange anisotropy (or dipolar
interaction) scales proportionally to the temperature $T$ in the
low temperature regime $T \ll J$ (but above the N\'{e}el or
spin-Peierls transition temperature). This, in fact,
appears to be observed in many quasi-one dimensional $S=1/2$
antiferromagnets\cite{Ajiro:CPC,Ohta,Demishev,Yamada:KCuF3,Yamada:CuGeO3,Yamada:NaVO}
including CPC, KCuF$_3$, CuGeO$_3$ and NaV$_2$O$_5$.
In the case of Cu benzoate\cite{Okuda}, 
there is a field-dependent diverging contribution to the linewidth
at low temperature due to a staggered field effect, as we will discuss
in Section~\ref{sec:stag}.
There seems to be another contribution to the linewidth,
which is  approximately $T$-linear and frequency-independent.
We presume the latter contribution is due to the exchange anisotropy.

\begin{figure}
\begin{center}
\epsfxsize=12cm
\epsfbox{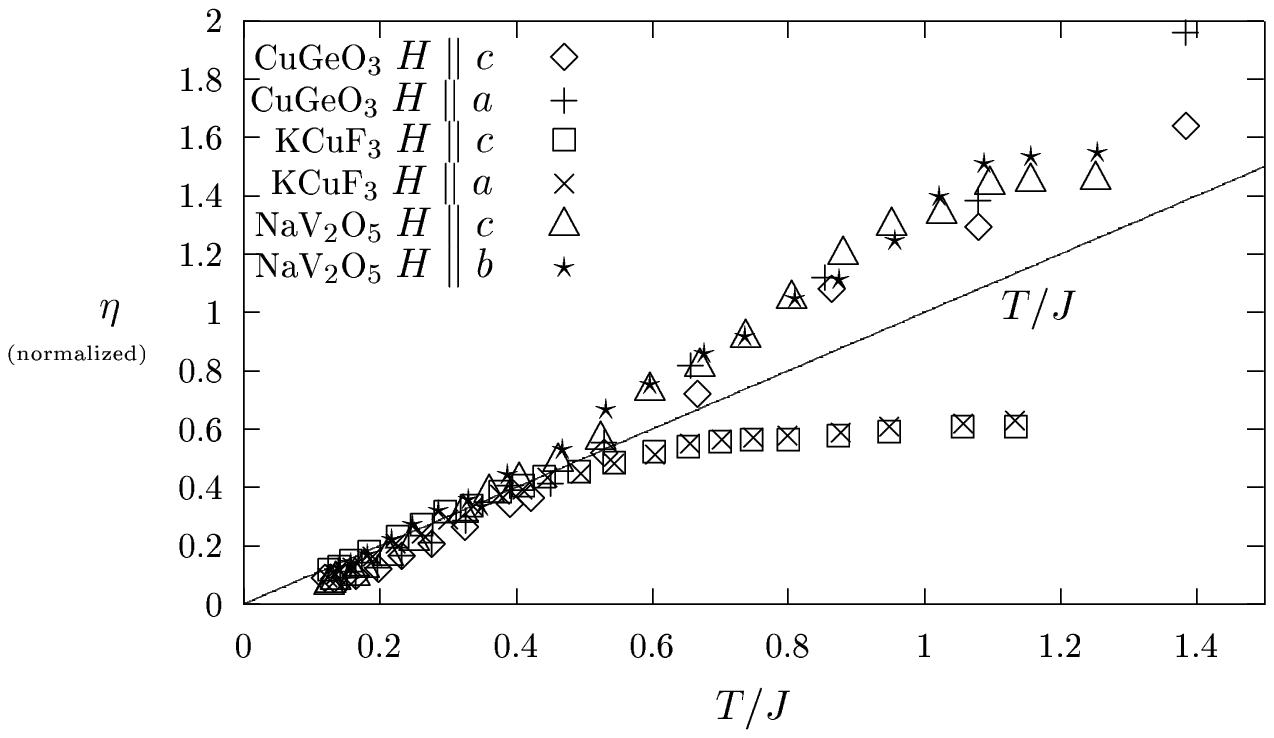}
\caption{
The temperature dependence of the ESR linewidth in
KCuF$_3$, CuGeO$_3$ and NaV$_2$O$_5$.
The data are taken from
Refs.~\protect\onlinecite{Yamada:KCuF3,Yamada:CuGeO3,Yamada:NaVO},
respectively.
The horizontal axis is the temperature $T$ normalized by the
exchange coupling $J$, and the
vertical axis is the normalized linewidth.
}
\label{fig:aniso}
\end{center}
\end{figure}

In Fig.~\ref{fig:aniso} we show the
observed\cite{Yamada:KCuF3,Yamada:CuGeO3,Yamada:NaVO}
ESR linewidth for KCuF$_3$, CuGeO$_3$ and NaV$_2$O$_5$,
as a function of the normalized temperature $T/J$.
We note that, these materials exhibit phase transitions
(such as N\'{e}el and spin-Peierls transitions)
at low enough temperatures, where the linewidth
appears to diverge.
Since we focus on one-dimensional systems in the present work,
in Fig.~\ref{fig:aniso} we have omitted such temperature regimes,
above which we may regard the system simply as a spin chain.
It could be possible that, however,
the displayed data are still affected by the interchain interactions,
the spin-Peierls instability etc.

An analysis on the linewidth in NaV$_2$O$_5$ similar to ours
was published previously by Zvyagin\cite{Zvyagin}.
However, we also remark that
the $T$-linear behavior of the linewidth due to an exchange
anisotropy was reported earlier in Ref.~\onlinecite{ESR-PRL}.
In fact, eq.~(3) in Ref.~\onlinecite{Zvyagin} is equivalent 
to eq.~(11) in Ref.~\onlinecite{ESR-PRL}.
Moreover, in Ref.~\onlinecite{Zvyagin}
it was argued that a bond-alternation perturbation
leads to a linewidth $\propto (J_1-J_2)^2/T^2$.
However, the argument (leading to Eq. (4) in Ref.~\onlinecite{Zvyagin})
cannot be correct per se, because the ESR
linewidth must remain strictly zero as long as all terms in the
Hamiltonian except the Zeeman term commute with the total spin operators,
as we reviewed in Sec.~\ref{sec:defesr}.
An isotropic bond-alternation has this property.
It is possible that an isotropic
bond-alternation perturbation $(J_1 - J_2)$
together with an anisotropic uniform exchange perturbation $\delta$
might lead to a width, but it would be suppressed by
the factor $\delta^2/J^2$ as the width should vanish when $\delta=0$.
In any case, a reliable derivation seems lacking so far.
We point out that the ESR spectrum cannot simply be related to the
boson propagator in the field theory, in the presence of a bond-alternation.
(See remarks below eq.~(\ref{eq:SpmJJ}).)

In Fig.~\ref{fig:aniso} we took $J=400$K, $150$K and $560$K
respectively\cite{Yamada:KCuF3,Yamada:CuGeO3,Yamada:NaVO}
for KCuF$_3$, CuGeO$_3$ and NaV$_2$O$_5$, while there are
some uncertainties in the estimate.
The linewidth is renormalized to be compared with $T/J$.
The low-temperature asymptotic behavior of the linewidth
indeed seems consistent, although not perfectly,
with the universal $T$-linear behavior we have derived.
On the other hand, it is difficult to discuss the predicted
logarithmic correction in the present data.
Regarding Fig.~\ref{fig:aniso} as a fitting, the low-temperature
asymptotic behavior reads
\begin{equation}
\frac{\eta}{T} \sim \left\{
\begin{array}{cc}
4.2 \times 10^{-4} & (\mbox{CuGeO$_3$}, H\parallel c) \\
4.7 \times 10^{-4} & (\mbox{CuGeO$_3$}, H\parallel a) \\
17 \times 10^{-4} & (\mbox{KCuF$_3$}, H\parallel c) \\
22 \times 10^{-4} & (\mbox{KCuF$_3$}, H\parallel a) \\
1.3 \times 10^{-4} & (\mbox{NaV$_2$O$_5$}, H\parallel c) \\
0.65 \times 10^{-4} & (\mbox{NaV$_2$O$_5$}, H\parallel b)
\end{array}
\right.  ,
\end{equation}
which are given as dimensionless numbers.
In these materials, the data for $H \parallel a$ and $H \parallel b$
are quite similar, and thus only one set of them is shown for
each material.

Comparing with our results~(\ref{eq:widthdeltaz}) and
(\ref{eq:widthdeltax}), the anisotropy $\delta/J$ seems to be
about a few percent.
It was argued\cite{Yamada:KCuF3,Yamada:CuGeO3,Yamada:NaVO} that,
in these material it is too (up to 10 times)
big compared to what we expect from Moriya's\cite{Moriya}
estimate $\delta \sim (\Delta g/g)^2 J$
where $\Delta g$ is the anisotropy of the $g$-tensor.
(Actually the discussion in
Refs.~\onlinecite{Yamada:KCuF3,Yamada:DM,Yamada:CuGeO3,Yamada:NaVO}
was based on the high-temperature limit.
See Section~\ref{sec:hight} for relation to our low-temperature theory.)
However, we believe that Moriya's formula is only valid as
an order-of-magnitude estimate. There is a room for a factor
which is presumably not too much different from $1$, but could still
allow the exchange anisotropy that is consistent with the observed
linewidth.

The linewidth deviates from the field theory result $\propto T$ at
higher temperatures. This is not surprising, since the field theory
is only valid in the low temperature $T \ll J$.
We will give more discussion on the crossover to the high-temperature
regime in Section~\ref{sec:hight}.
On the other hand, if all the materials can be regarded as
standard Heisenberg antiferromagnetic chains with the same type
of anisotropy,
we would expect the linewidth to be a universal function of $T/J$.
However, in Fig.~\ref{fig:aniso}
it is evident that the linewidth behaves differently at high
temperature, especially in KCuF$_3$.
This suggests that not all of them can be described
by the standard Heisenberg antiferromagnetic chain~(\ref{eq:HAFC})
with the same type of anisotropy.
We remark that the low-temperature asymptotic behavior
should be universal for a certain class of Hamiltonians, but
the explicit coefficients obtained in eqs.~(\ref{eq:widthdeltaz}) and
(\ref{eq:widthdeltax}) are specific to the standard
Hamiltonian~(\ref{eq:HAFC}).

Certainly, there are many questions still to be understood.
An important problem is the dependence on the direction of the
applied field.
In the case of NaV$_2$O$_5$, the observed linewidth at low temperature
is twice as large when $H \parallel c$ compared as when $H \perp c$.
This is consistent with our result, if
an exchange anisotropy with the single anisotropy axis parallel to $c$
is assumed.
However, in the case of CuGeO$_3$ and KCuF$_3$,
the observed linewidth for
$H \parallel c$ is smaller than that for $H\parallel a$ and $H\parallel b$.
This kind of angular dependence cannot
be explained with an exchange anisotropy with a single anisotropy
axis. This suggests that we have to consider more general 
types of anisotropy, or some other effects.

A complete theoretical description of the experimental data in these
materials is left for the future.
Nevertheless, we believe that the universal
decrease of ESR linewidth at low temperatures
in $S=1/2$ antiferromagnetic chains
is basically understood with our theory.
Ours is presumably the first\cite{ESR-PRL} microscopic derivation
of this approximately $T$-linear linewidth.
In Refs.~\onlinecite{Yamada:KCuF3,Yamada:DM,Yamada:CuGeO3,Yamada:NaVO}
a completely different interpretation was proposed.
However, we will argue against it in Section~\ref{sec:hight}.

\section{ESR in an XXZ antiferromagnet}
\label{sec:xxz}

So far in this paper, we have restricted ourselves to the
case of small anisotropy.
However, in principle ESR can be measured in a system
which is far from isotropic.
To apply the self-energy formalism to
a not small anisotropy, one has to sum up higher orders
of the perturbation.
In addition, the foundation of our
self-energy formalism
based on the weakly broken $SU(2)$ symmetry
may be questionable in such cases,
because the $SU(2)$ symmetry is strongly
broken in the spin Hamiltonian.

However, there is one case in which we can study ESR with a strong
anisotropy: an easy-plane XXZ antiferromagnet with a field applied
perpendicular to the easy plane. This is nothing but the isotropic
Heisenberg antiferromagnet with a negative exchange anisotropy
parallel to the applied field~(\ref{eq:XXZ}), with $\delta < 0$.
Here we can apply the direct calculation introduced in
Sec.~\ref{sec:anis-z}.

The compactification radius $R$ for the XXZ model with a given
anisotropy $\delta$ is known from Bethe Ansatz exact solution and
is given  in eq.~(\ref{eq:Rofdelta}). Using this radius, the ESR
absorption spectrum given by the Green's
function~(\ref{eq:Schulz}) of the vertex operator with the
conformal weight~(\ref{eq:Delta}),(\ref{eq:Delta-R}). Since
$\delta$ is not small, the spectrum is no longer a simple
Lorentzian, except at low enough temperature $T \ll H \ll J$ where
the spectrum reduces to the Lorentzian~(\ref{eq:Gamma-Lorentz}).

In this Lorentzian case, the width here does {\em not} reduce to
the previous one~(\ref{eq:widthdeltaz}) which was proportional to
$\delta^2$, even in the limit $\delta \rightarrow 0 $. The reason
of this disagreement is that they describe different regimes. The
result~(\ref{eq:widthdeltaz}) is valid when the energy scale ${\rm
max}(T,H)$ is above the crossover energy $E_c = e^{-1/\epsilon}$,
while the present result is valid if the relevant energy scale $T$
and $H$ are both below $E_c$. For a small anisotropy, the
crossover scale is exponentially small, making
eq.~(\ref{eq:widthdeltaz}) realistic for the experimentally
accessible regime.

For a small exchange anisotropy and above the crossover energy $E_c$,
the width is proportional to $\lambda^2$ in the leading order of
perturbation theory; the width is insensitive to the sign of the
anisotropy (easy-plane or easy-axis). However, when the anisotropy
is large or $T, H \ll E_c$, this symmetry no longer holds. In
fact, the system in the zero temperature limit is gapless for an
easy-plane anisotropy ($\delta < 0$) while it acquires a gap $\sim
E_c$ for an easy-axis anisotropy ($\delta > 0$). In the gapful
case $\delta > 0$ and $T,H \ll E_c$,  ESR probes the creation of
the elementary excitation above the groundstate; the absorption
spectrum then has a sharp peak centered at the energy of the gap.

\section{Transverse staggered field}
\label{sec:stag}

As we have discussed in Section~\ref{sec:perturbations}, a
staggered field is the most relevant perturbation of the isotropic
Heisenberg antiferromagnet. Breaking the $SU(2)$ symmetry, the
staggered field affects also the ESR spectrum. Here we discuss the
effect by the field theory methods described in previous sections,
and then explain the mysterious observations in ESR
experiments~\cite{Okuda,Oshima:AFMR} on Cu Benzoate in the 1970's
which were recently confirmed and extended \cite{Asano}.

Let us focus on the case of a transverse staggered field
\begin{equation}
   \calH' = h \sum_j (-1)^j S^x_j .
\label{eq:trstag}
\end{equation}
As we have discussed already, the staggered field is mapped to the
operator
\begin{equation}
    n^x \sim k \cos{(2 \pi R \tilde{\phi})}
\label{eq:nxcosine}
\end{equation}
which has scaling dimension $1/2$.
A standard scaling analysis similar to that in
Section~\ref{sec:anis-z} shows that,
ignoring the logarithmic correction,
the linewidth should be given as
\begin{equation}
 \eta = T g( \frac{E_g}{T}, \frac{H}{T}),
\end{equation}
where $E_g$ is the excitation gap\cite{Cubenz-PRL,Cubenz-PRB} due to the
staggered field  proportional to $h^{2/3} J^{1/3}$.
Again, the scaling argument alone cannot determine the actual
form of the scaling function $g$.

\subsection{Self-energy approach}

As we have discussed,
The staggered transverse field~(\ref{eq:trstag}) is mapped
to the field theory operator $n^x$:
\begin{equation}
    \calH' = h \sum_j (-1)^j S^x_j \sim k h \int n^x(r) dr
\label{eq:trstagn}
\end{equation}
where $k$ is a constant,
and we normalize $n^x$ by $\langle n^x(r) n^x(0) \rangle = 1/r$.
Namely, $k^2$ gives the correlation amplitude
$\langle S^x_0 S^x_j \rangle \sim (-1)^j k^2/j$.
This form is not affected by the application of the
magnetic field $H$, except for the possible renormalization
of the amplitude $k$ and the exponent, which we will ignore.

The $SU(2)$ WZW field theory with the perturbation $n^x$ has
rotational symmetry about the $x$-axis. While the original spin
problem is not invariant under a rotation about the $x$-axis due
to the applied field, the effective field theory does have this
symmetry. As a consequence, correlation functions of the type
$\langle J^x J^y \rangle$ vanishes. Thus we can apply the
self-energy method by reducing the ESR spectrum to Green's
function of the bosonic field, as discussed in
Section~\ref{sec:self-energy}.

The transverse staggered field in the $x$ direction breaks the
rotational symmetry in the $xy$-plane, leading to polarization
dependence. Calculations similar to those in
Section~\ref{sec:anis-x} lead to the same
result~(\ref{eq:GRxxex}),(\ref{eq:GRyyex}) and (\ref{eq:GRxyex}).
The polarization dependence is again consistent with the rigorous
relation~(\ref{eq:pol-dependence}) which can be applied to the
present case.

The self-energy $\Pi$ is now replaced by the boson self-energy
in the presence of the perturbation $kh \cos{\sqrt{2\pi}\phi}$.
Again, arguments similar to those in Section~\ref{sec:anis-x}
can be applied to obtain the result
\begin{equation}
\Pi^R (\omega,q) =
 2 \pi (kh)^2 [ G^R_{(1/4,1/4)}(\omega,q) - G^R_{(1/4,1/4)}(0,0) ],
\end{equation}
where the second term comes from the tadpole term.

The self-energy is a smooth function of $\omega$ near the resonance
$\omega \sim H$.
Thus, the lineshape is Lorentzian near the center of the resonance,
with the width and shift determined by the self-energy at
$(\omega, q) = (H,H)$.
The imaginary part of the second, tadpole term vanishes according to
eq.~(\ref{eq:Schulz}).
Using eq.~(\ref{eq:selfwidth}), the linewidth is given by
\begin{equation}
\eta = \frac{\pi k^2 h^2}{H}
        {\rm Im} \left[  - G^R_{(1/4,1/4)}(H,H)  \right].
\label{eq:selfstagwidth}
\end{equation}

{F}rom eq.~(\ref{eq:Schulz}), the linewidth shows quite a nontrivial
dependence on the applied field $H$ and temperature $T$. However,
in the weak field regime $H \ll T$, the formula can be simplified
and linewidth has simple $T^{-2}$ dependence on the temperature.
\begin{equation}
\eta = \frac{\pi k^2}{4}
\left( \frac{\Gamma(\frac{1}{4})}{\Gamma(\frac{3}{4})} \right)^2
\frac{h^2}{T^2} .
\end{equation}
We note that this is consistent with the scaling analysis. (Recall
that we have set $v=1$.)

The correlation amplitude $k^2$ was recently determined exactly
for the $S=1/2$ Heisenberg
antiferromagnet\cite{Lukyanov,Affleck:log,Lukyanov:amp}
with a logarithmic correction due to the presence of
marginal operators:
\begin{equation}
(-1)^r \langle S^z(r) S^z(0) \rangle
     \sim \frac{1}{(2 \pi)^{3/2}} \frac{\ln{r}}{r} .
\end{equation}
The logarithmic correction is translated into a $\ln{(J/T)}$
factor in the ESR, where the temperature gives the IR cutoff. Thus
we obtain (upon reinstating $v=\pi J/2$)
\begin{equation}
\eta = \frac{1}{16} \sqrt{ \frac{\pi}{2} }
    \left( \frac{\Gamma(\frac{1}{4})}{\Gamma(\frac{3}{4})} \right)^2
    \frac{J h^2}{T^2}  \ln{(\frac{J}{T})}
\sim 0.685701   \frac{J h^2}{T^2}  \ln{(\frac{J}{T})}.
\label{eq:Hstagwidth}
\end{equation}
Implication of this result on the experiments will be discussed
in Sec.~\ref{sec:stagexp}.

\subsection{MK approach}
\label{sec:stagMK}

Since the lineshape is Lorentzian, the MK formula should be
valid also in this case, provided the correlation function
is evaluated appropriately.
There are two ways to evaluate the commutator~(\ref{eq:defA})
appearing in the MK formula: to take the continuum
limit before calculating the commutator, or to first calculate the
commutator in terms of the original spin variable and then take the continuum
limit.
We think the former is generally more reliable,
since the field theory only deals with universal low-energy phenomena
while the Lorentzian assumption of MK formula would be valid at
best in the long-time limit.
In the present case, the two methods give
the same result as we will show below.

Taking the continuum limit first, we calculate the commutator
between the field theory operators~(\ref{eq:trstagn})
and~(\ref{eq:S-opsN}). The standard relation between the
commutator and OPE leads to
\begin{eqnarray}
\calA &=& [ \calH' , S^- ]
\nonumber \\
&=& [ h k  \int n^x(r) dr, \frac{1}{8 \pi^2} \int J^-_R(r) e^{-i H r}
+J^-_L(r) e^{i H r}]\nonumber \\
&=&  (hk/2)
\int [e^{-i\sqrt{2\pi}\phi} e^{-i H r}+e^{i \sqrt{2\pi} \phi} e^{i H r}] dr
\label{eq:trstagA}
\end{eqnarray}
On the other hand, in the original spin representation,
the commutator is easily evaluated as
\begin{equation}
 \calA = [ \calH' , S^- ] = h \sum_j (-1)^j S^z_j ,
\label{eq:calAstag}
\end{equation}
namely the longitudinal staggered field with the coefficient $h$.
Taking continuum limit, it agrees with~(\ref{eq:trstagA}).

Thus, from the MK formula, the linewidth is given by
\begin{equation}
\eta = \frac{k^2 h^2}{2 \chi_u H}
    {\rm Im} \left[  - G^R_{(1/4,1/4)}(H,H)  \right] .
\label{eq:trstagMK}
\end{equation}
Again using eq.~(\ref{eq:chiu}),
this agrees exactly with the result~(\ref{eq:selfstagwidth})
obtained in the self-energy approach.

\subsection{Shift of the resonance frequency}
\label{sec:stagshift}

In the present case, there is no shift to first order in $h$. In
fact, the first term in the MK formula (\ref{eq:MKshift})
vanishes in the present case. The lowest order shift is thus
second order in $h$. This is given by either the  MK
formula~(\ref{eq:MKshift}) or by ${\rm Re}\Pi^R(H,H)/(2H)$ in the
self-energy approach. Again, both approaches give the same result
for the frequency shift:
\begin{equation}
\Delta \omega =
\eta = \frac{\pi k^2 h^2}{H}
{\rm Re} \left[ - G^R_{(1/4,1/4)}(0,0) + G^R_{(1/4,1/4)}(H,H)  \right] .
\label{eq:stagshift}
\end{equation}

This is a straightforward consequence of the self-energy approach.
On the other hand, the derivation from the MK approach might need
an explanation. While the second term proportional to $-
G^R_{(1/4,1/4)}(H,H)$ just comes from $G^R_{\calA \calAdag}$ in
the MK formula~(\ref{eq:MKshift}), the first term (proportional to
$- G^R_{(1/4,1/4)}(0,0)$) is less obvious. From
eq.~(\ref{eq:calAstag}), the commutator in the first term of the
MK formula~(\ref{eq:MKshift}) is given by
\begin{equation}
   [ \calA, S^- ] =  h \sum_j (-1)^j S^-
\end{equation}
Its expectation value vanishes if evaluated in the absence of
the staggered field $\calH'$.
However, taking the staggered field perturbation into account,
\begin{equation}
   \langle [ \calA, S^- ] \rangle =  - h^2 \chi_s + O(h^3),
\end{equation}
where $\chi_s$ is the (transverse) staggered susceptibility.
By the linear response theory, we have
\begin{equation}
   \chi_s = - k^2 {\rm Re} G^R_{(1/4,1/4)}(0,0),
\end{equation}
which leads to~(\ref{eq:stagshift}), with the replacement
of $\chi_u$ by its zero-temperature limit~(\ref{eq:chiu}).

For the standard $S=1/2$ Heisenberg antiferromagnetic chain,
we can apply the exact result on the correlation amplitude
as we did for the width.
We obtain
\begin{equation}
\Delta \omega =
    \frac{1}{8}\sqrt{\pi \over 2} \ln{(\frac{J}{T})}
\frac{J h^2}{HT}
\left(\frac{\Gamma(\frac{1}{4})}{\Gamma(\frac{3}{4})}\right)^2
\left[ 1 - \frac{\Gamma(\frac{3}{4})}{\Gamma(\frac{1}{4})}
        {\rm Re} \left\{
{\Gamma(\frac{1}{4}- i \frac{H}{2 \pi T})\over
 \Gamma(\frac{3}{4}- i \frac{H}{2 \pi T})}
\right\}
\right]
\label{eq:Hstagshift1}
\end{equation}
For small field $H$ compared to temperature $T$, we obtain
by Taylor expansion of the Gamma function
\begin{equation}
\Delta \omega = .344057 \frac{J h^2 H}{T^3}\ln {(\frac{J}{T})}.
\label{eq:Hstagshift2}
\end{equation}
(An incorrect prefactor was given in Ref.~\onlinecite{ESR-PRL}.)
Namely, we obtain the positive shift which rapidly increases
with decreasing temperature.

The shift in the presence of the staggered $g$-tensor
was previously discussed by Nagata\cite{Nagata} using
the formula
\begin{equation}
\Delta \omega = - \frac{1}{2 \chi_u H} \langle [[S^+, \calH'],S^-] \rangle .
\label{eq:NT}
\end{equation}
derived in Refs.~\onlinecite{KanamoriTachiki,NT}. (See also Appendix.)
In the present case, it is reduced to the expectation value
of the staggered field $h \sum_j (-1)^j S^x_j$.
The leading order of the shift in the perturbation $h$ is thus given by
\begin{equation}
\Delta \omega = \frac{h^2}{2 \chi_u H} \chi_s,
\label{eq:NTstag}
\end{equation}
where $\chi_s$ is the staggered susceptibility~\cite{Nagata}. The
positive frequency shift (ie. negative field shift) was argued to
be consistent with the experiment\cite{Oshima:gshift,Nagata}.
On the other hand,
the theoretical result in Ref.~\onlinecite{Nagata}
is not in  quantitative agreement,
partly due to the evaluation of $\chi_s$ in the high-temperature
classical limit.
However, we believe that eq.~(\ref{eq:NTstag}) itself
is not quite correct even if $\chi_s$ were evaluated exactly.
In fact, eq.~(\ref{eq:NTstag}) differs
from ours~(\ref{eq:stagshift}).
Interestingly, eq.~(\ref{eq:NTstag}) is
equivalent to including only the tadpole contribution in the
self-energy approach. The discrepancy becomes particularly
important at low magnetic field. In the limit of $H \rightarrow 0$
at fixed $h$ and $T$ (although this limit is not realistic in
experiment) the MK/self-energy approach predict the shift linear
in $H$ but (\ref{eq:NTstag}) gives a diverging shift $\sim 1/H$,
which is presumably unphysical. While eq.~(\ref{eq:NT}) captures some
physics of the frequency shift, it fails to include more subtle
effects of fluctuation, presumably because of the oversimplified
ansatz and of not including the long-timescale dynamics.

\subsection{Comparison with the experiments}
\label{sec:stagexp}

The result~(\ref{eq:Hstagwidth}) of the perturbation theory
implies an interesting behavior of the ESR linewidth in
materials such as Cu benzoate.
As we have discussed in Section~\ref{sec:stagfield}, there
an effective transverse staggered field is induced proportionally
to the applied field ($h = cH$), and the proportionality constant $c$
depends strongly on the direction of the applied field.
Thus, the linewidth increases as $\propto T^{-2}$ as the temperature
is lowered. Furthermore, it depends on the applied field (or the resonance
frequency) $H$ and on the direction of the applied field.
This very characteristic behavior is not expected for the exchange
anisotropy.
In fact, these features were actually observed\cite{Okuda} nearly 30 years ago
in ESR on Cu benzoate and apparently have not been understood until
recently. Our results give a natural understanding of these
observations\cite{ESR-PRL}.

The only unknown parameters in Cu benzoate were two components of
DM vector.
We have chosen\cite{ESR-PRL}
\begin{equation} 
(D_{a''},D_{c''}) = (0.13,0.02)J
\label{eq:DMfit}
\end{equation}
which seemed most reasonable to fit ESR data\cite{Okuda}.
It is also roughly consistent with other experiments\cite{Cubenz-PRB}
such as neutron scattering, although not perfectly.
This choice of DM vector fit rather nicely the direction dependence
(Fig.~1 of Ref.~\onlinecite{ESR-PRL}), temperature and field dependence
(Fig.~2 of Ref.~\onlinecite{ESR-PRL}).
However, we should note that
the determination of the logarithmic
correction in a practical fitting is a difficult problem; the leading
log correction is only valid in the low temperature limit.
Our fittings were done setting the logarithmic factor to unity.

\begin{figure}
\begin{center}
\epsfxsize=12cm
\epsfbox{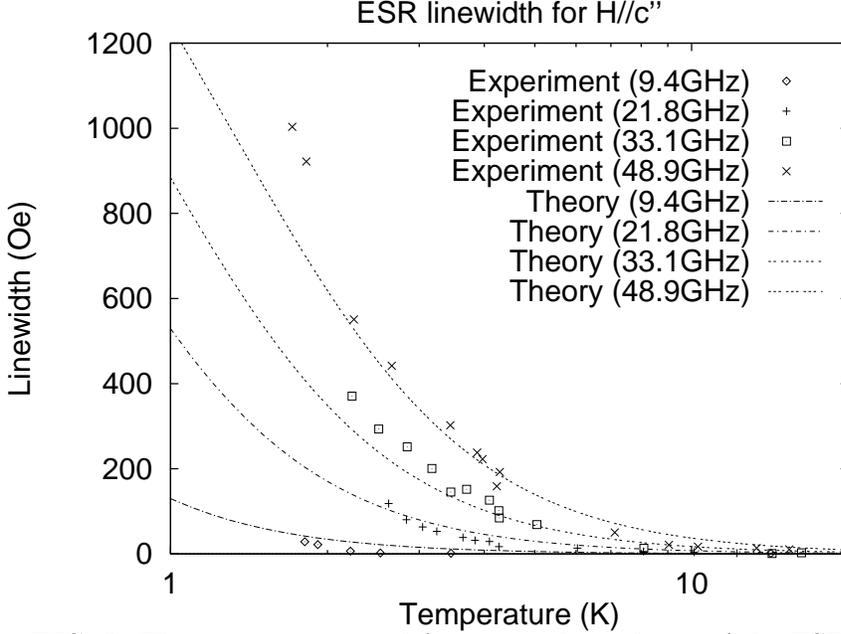}
\caption{
The temperature and frequency dependence of the ESR linewidth
for $H \parallel c''$~\protect\onlinecite{Okuda},
after subtracting the frequency independent part.
It is compared our theory~(\ref{eq:Hstagwidth}).
See also Fig.~2 of Ref.~\protect\onlinecite{ESR-PRL} for the
same comparison without the subtraction.
}
\label{fig:width-subtr}
\end{center}
\end{figure}

There is some discrepancy between the theory and the experiments.
We see, in the experimental data, a field(frequency)-independent contribution
which appears to be approximately linear in temperature.
This is presumably due to effects other than the staggered field.
A probable mechanism is the effect of an exchange anisotropy, which
gives a linewidth which is $T$-linear and independent of the field.
If we subtract the field-independent contribution from the
experimental data at the price of introducing additional fitting
parameters, the agreement becomes better as shown in
Fig.~\ref{fig:width-subtr}.

Recently Asano et al. made a detailed experimental study\cite{Asano}
on ESR in Cu benzoate.
They also confirmed our prediction on the linewidth at higher field.
In addition,
they found that, when the temperature is small compared to $J$ but
not too low,
the shift is  consistent with
our prediction~(\ref{eq:Hstagshift2}).
See Fig.~2 of Ref.~\onlinecite{Asano}.
Moreover, we can read off the proportionality constant from their Fig.~2
as
\begin{equation}
        \Delta \omega \sim 0.053 (\frac{H}{T})^3
\end{equation}
for $H\parallel c$,
where $\Delta \omega$ and $H$ are measured in Tesla while $T$ is
in Kelvin.
On the other hand, using the DM vector~(\ref{eq:DMfit}),
e find that $h = 0.095 H$ for $H \parallel c$.
Combining this with~(\ref{eq:Hstagshift2}), we the theoretical prediction
\begin{equation}
        \Delta \omega \sim 0.042 (\frac{H}{T})^3 ,
\end{equation}
where we have again replaced the logarithm $\ln{J/T}$ with unity.
Considering the subtlety of the logarithmic correction, the
agreement between the theory and the experiment is rather good.

Thus our perturbative results agrees well with the experiments.
However, at very low temperatures,
the lineshape evolves differently than what we expect from the
lowest order perturbation theory.
This will be discussed in the next subsection.

\subsection{Resonance at very low temperature}
\label{sec:vltemp}

So far, our analysis was perturbative in the staggered field $h$.
While the perturbation theory seems reasonable for a small
staggered field, it eventually fails at lower temperature
where the effect of the staggered field is enhanced.
In fact, the perturbative expansion turns out to be
an expansion in $Jh^2/T^3$, which is divergent at low enough
temperature.

The effective field theory describing the $S=1/2$ Heisenberg
antiferromagnetic chain with a staggered field
is given by~(\ref{eq:freebosonL}) perturbed with~(\ref{eq:nxcosine}).
As discussed in Ref.~\onlinecite{Cubenz-PRL},
this is nothing but the sine-Gordon field theory, which is
one of the best understood strongly interacting field theories.
Since the interaction term~(\ref{eq:nxcosine}) is relevant,
the sine-Gordon field theory is massive, ie. has a finite
excitation gap $E_g$ above the groundstate.
The elementary excitations of the sine-Gordon model consist
of solitons, antisolitons and breathers which are boundstates
of a soliton and an antisoliton.

The perturbation theory is expected to be valid only for $T \gg
E_g$. Here we consider the opposite limit $T \ll E_g$, where the
system is essentially in the groundstate. Then we obtain quite a
different picture. It is still valid that the ESR spectrum is
given by the $\langle \phi \phi \rangle$  Green's function at frequency and
momentum $H$. However, we have to consider the zero-temperature
Green's function in a non-perturbative way.

In the present case ($\beta = \sqrt{2 \pi}$), the lowest
excitations are 1st breather, soliton and antisoliton, which
form an $SU(2)$ triplet.
Thus the excitation gap $E_g$ is identical to the
first breather mass $M_1$.
The boson field $\phi$ couples to the 1st breather, and thus
its propagator is given by
\begin{equation}
    \langle \phi \phi \rangle (\omega,q) \sim
        \frac{Z^{\phi}}{\omega^2 - q^2 - {M_1}^2}.
\end{equation}
where $Z^{\phi}$ is the wavefunction renormalization constant
obtained exactly\cite{Karowski} as
\begin{equation}
Z^{\phi} = (1+ \nu)
\frac{\frac{\pi \nu}{2}}{2 \sin{\frac{\pi \nu}{2}}}
\exp{\left( - \frac{1}{\pi} \int_0^{\pi \nu} \frac{t}{\sin{t}} dt \right)},
\end{equation}
where $\nu = \beta^2/(8 \pi - \beta^2)$.
For the present case, $\nu =1/3$ and thus $Z^{\phi} = 0.978689$.

{F}rom this, we immediately find that the ESR spectrum at zero
temperature is given by a delta-function
\begin{equation}
    - {\rm Im} G^R_{S^+S^-}(\omega) \approx
 {\left( H+\sqrt{H^2+M_1^2}\right)^2\over 2
\sqrt{H^2+M_1^2}}
  \delta(\omega - \sqrt{H^2 + {M_1}^2}) .
\end{equation}
Since the wavefunction renormalization $Z^{\phi}$ is close to
unity, the intensity is identical to that of a free resonance.

Thus we obtain a rather complicated behavior of ESR in the
presence of the staggered field.
As the temperature is lowered, the linewidth increases
in the perturbative regime ($T \gg E_g$) as we discussed,
but at lower temperature ($T \ll E_g$) we see a revival
of a sharp resonance.
The width of the resonance vanishes at zero temperature.
At small but finite temperature $0 < T \ll E_g$,
the resonance may be broadened due to
the thermally activated excitations, but presumably the effect is
only of the order of the density of such excitations $\sim \exp{(- E_g/T)}$.

On the other hand, the ESR frequency at zero (or very low)
temperature does receive a shift due to the staggered field.
Namely, the resonance frequency $\omega$ is given by
\begin{equation}
 \omega = \sqrt{H^2 + {M_1}^2} ,
\label{eq:bresonance}
\end{equation}
compared to the Zeeman frequency $H$.
For small mass $M_1 \ll H$, the shift is given as
\begin{equation}
    \Delta \omega \sim \frac{{M_1}^2}{2H}
        \sim 1.57878 \left(\ln{\frac{J}{h}}\right)^{1/3}
        \frac{J^{2/3} h^{4/3}}{H},
\label{eq:brshift}
\end{equation}
where we used the result of the breather mass (field-induced gap)
$E_g = M_1 \sim 1.77695 [\ln{(J/h)}]^{1/6} (J h^2)^{1/3}$ in
Refs.~\onlinecite{Cubenz-PRL,Cubenz-PRB}. Here we emphasize that our
self-energy approach is valid beyond the lowest order of
perturbation theory, unlike the MK formula. Thus it allows us a
non-perturbative analysis such as the above.

Our prediction agrees quite well with the experimental result in
Ref.~\onlinecite{Oshima:AFMR}, as discussed in
Ref.~\onlinecite{ESR-PRL}.
The non-trivial evolution of the
lineshape was indeed observed in the experiment in 1970s.
Moreover, using the same parameter~(\ref{eq:DMfit}) we have used
for the perturbative analysis,
we are able to reproduce the direction dependence of the resonance
frequency at very low temperature quite well with
eq.~(\ref{eq:bresonance}) as shown in Fig. 3 of
Ref.~\onlinecite{ESR-PRL}.
However, the data were only shown at fixed temperature and fixed frequency in
Ref.~\onlinecite{Oshima:AFMR}. Thus several other predictions of
our theory could not be compared. After our
proposal\cite{ESR-PRL}, Asano et al. studied\cite{Asano}  ESR in
Cu benzoate at low temperature and at higher field. They confirmed
the crossover to the non-perturbative regime, and that the
resonance at very low temperature agreed with the
prediction~(\ref{eq:brshift}) for various fields. Moreover, the
crossover between the perturbative and the non-perturbative regime
occurs at temperature $T \sim E_g$, consistently with our picture.
The broadening at the non-perturbative regime was also consistent
with the picture $e^{-E_g/T}$

On the other hand, the precise lineshape at the crossover
temperature regime $T \sim E_g$ requires a non-perturbative
calculation of the correlation function of the boson field in the
sine-Gordon field theory {\em at finite temperature}. Despite
remarkably many exact results on the theory based on the
integrability, calculation of the finite temperature correlation
remains an unsolved problem. The ESR lineshape in Cu benzoate
provides a set of rather precise experimental data for the finite
temperature correlation function in the sine-Gordon field theory.
It is hoped that future theoretical progress will enable us to
compare theoretical non-perturbative result with the ESR data
in the crossover temperature regime $T \sim E_g$.

In Ref.~\onlinecite{Oshima:AFMR}, the sharp resonance at very low
temperature is considered to be the ``antiferromagnetic
resonance,'' which reflects the N\'{e}el ordering due to the
interchain interaction. In particular, they identified the
appearance of the sharp resonance at very low temperature as the
N\'{e}el transition. However, a recent $\mu$SR experiment on Cu benzoate
reveals\cite{Nojiri} that a N\'{e}el ordering does not occur even
down to $20$mK. We believe that the evolution of the ESR lineshape
is primarily explained within our one-dimensional theory taking
the effective staggered field into account. On the other hand, we
also note that the interpretation in Ref.~\onlinecite{Oshima:AFMR} is not
totally different from ours; the system has a long-range magnetic
order in both theories. The difference is that the order is
induced {\em spontaneously} due to the interchain interaction in
Ref.~\onlinecite{Oshima:AFMR} while it is forced externally by the
staggered field in our picture. Spin-wave theory can also be 
applied to the externally ordered state; the resonance at very low
temperature would be then identified
with the ``upper mode'' $E_+$ (see
eq.~(3.12) of Ref.~\onlinecite{Cubenz-PRB}) which has a
qualitatively similar dependence on $h$ and $H$ to
eq.~(\ref{eq:bresonance}).
(We thank H. Shiba for pointing this out.)
Quantitatively, however, the
sine-Gordon field theory is expected to work better for a small
staggered field $h$.

\section{ESR at higher temperatures}
\label{sec:hight}

In this paper, we have developed a field-theory approach to ESR in
quantum spin chains. The field theory is a low-energy effective
theory, and is only valid at low temperatures compared to the
exchange coupling. Here we would like to consider ESR in the other
extreme, namely the high temperature limit using Kubo-Tomita
theory. We will also discuss the crossover between the
low-temperature and high-temperature regime.

\subsection{Exchange anisotropy}
\label{sec:hight-anis}

\begin{figure}
\begin{center}
\epsfxsize=8cm
\epsfbox{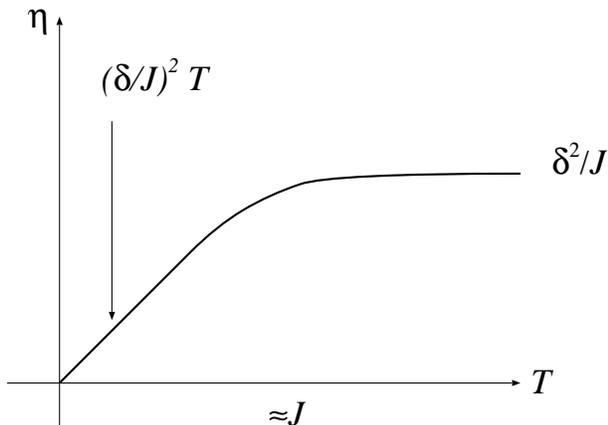}
\caption{
Simplest scenario of the temperature dependence of the linewidth
in the presence of an exchange anisotropy.
The $T$-linear behavior at low temperature predicted by the
field theory method crossovers smoothly
to the constant $\sim \delta^2/J$ predicted by the Kubo-Tomita
theory at high temperature limit.
The crossover takes place at $T\sim J$, which is the limit
of the validity of the field theory approach.}
\label{fig:anis-cross}
\end{center}
\end{figure}

For the exchange anisotropy~(\ref{eq:anis}) in a generic
direction, the KT formula~(\ref{eq:KTwidth}) has been
applied to the linewidth in the literature.
The result is
\begin{equation}
    \eta \propto \frac{\delta^2}{J},
\end{equation}
where we have ignored the direction dependence.
It is difficult to discuss the intermediate temperature
regime either with the existing theories or with our field theory
approach.
However, our result can be naturally related to the
high-temperature limit, assuming a smooth crossover at temperature
$T \sim J$, namely if the $T$-linear
behavior~(\ref{eq:zzwidth}),(\ref{eq:xxwidth})
is cut off at $T \sim J$ as shown in
Fig.~\ref{fig:anis-cross}.
In fact, this simple scenario seems to agree with
the experimental
results\cite{Ajiro:CPC,Ohta,Demishev,Yamada:KCuF3,Yamada:CuGeO3,Yamada:NaVO}
on CPC, KCuF$_3$, CuGeO$_3$ and NaV$_2$O$_5$,
which we think the exchange anisotropy (including the dipolar
interaction) is the primary mechanism of the broadening.
See Fig.~\ref{fig:aniso} for some of the examples.

We note that, while the low-temperature asymptotic behavior described
by the field theory is universal, the crossover to the high-temperature
regime is expected to be non-universal.
The linewidth as a function of the temperature would
depend, for example, on the next-nearest-neighbor interaction
introduced additionally to the standard Hamiltonian~(\ref{eq:HAFC}).

\subsection{Staggered field}

\begin{figure}
\begin{center}
\epsfxsize=8cm
\epsfbox{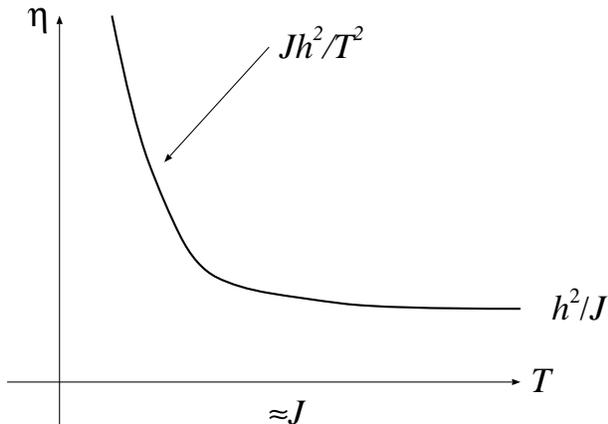}
\caption{
Simplest scenario of the temperature dependence of the linewidth
in the presence of a staggered field, interpolating the
low-temperature field theory result $Jh^2/T^2$ and the
high-temperature result $h^2/J$.}
\label{fig:stag-cross}
\end{center}
\end{figure}

No literature on the effect of a staggered field in ESR linewidth
is known to us.
The application of the KT formula~(\ref{eq:KTwidth}) to the
staggered field perturbation~(\ref{eq:trstag}) is nevertheless
straightforward, giving the linewidth
\begin{equation}
    \eta \propto \frac{h^2}{J},
\end{equation}
at the high temperature limit.
This is again consistent with the low-temperature field theory
result eq.~(\ref{eq:Hstagwidth}), assuming a smooth crossover at $T
\sim J$, as shown in Fig.~\ref{fig:stag-cross}.

\subsection{Dzyaloshinskii-Moriya interaction}
\label{sec:hight-DM}

While the effects of a DM interaction~(\ref{eq:DMterm}) has been
discussed in the literature\cite{Date,Yamada:KCuF3,Yamada:DM},
we believe there is
a rather serious problem with these previous treatments. A direct
application of the KT formula~(\ref{eq:KTwidth}), as was made
previously, yields the width
\begin{equation}
    \eta \propto \frac{D^2}{J},
\label{eq:etaDMdirect}
\end{equation}
where we have again ignored the angle dependence.

On the other hand, a staggered DM interaction can be reduced to an
exchange interaction $\delta \sim D^2/J$ and a transverse
staggered field $h \sim DH/J$ via an exact transformation
discussed in Sec.~\ref{sec:stagfield}.
If we apply the KT formula after the transformation, we obtain
\begin{equation}
    \eta \propto \frac{D^4}{J^3} + \frac{D^2 H^2}{J^3},
\label{eq:etaDMelim}
\end{equation}
where we ignored constants of $O(1)$. This actually differs
substantially from the result of the direct
application~(\ref{eq:etaDMdirect}). In a typical situation, $D/J
\sim 0.1$ and $H \ll D < J$ so that $\eta \sim 0.01 J$ from
eq.~(\ref{eq:etaDMdirect}) while $\eta \sim 10^{-4} J$ from
eq.~(\ref{eq:etaDMelim}), which means a factor of $100$
difference.
(Actually we have to know the numerical coefficient, which has been
ignored so far, in order to discuss the absolute value of the width.)
In addition, it is argued\cite{KSEA} that there exists an
exchange anisotropy (before the transformation) which accompanies
the DM interaction, and cancels the anisotropy coming from the DM
interaction. The discrepancy would be even greater when this
happens.

Obviously, both results cannot be true at the same time
(while they could be both wrong.)
What we believe is that the latter approach eliminating
the DM interaction first is appropriate, and the direct
application of the KT formula to the DM interaction
is incorrect.
A possible reason why the direct application~(\ref{eq:etaDMdirect})
fails is as follows.
In the latter approach based on the transformation~(\ref{eq:etaDMelim}),
the physical total spin operator $S^{x,y}$ is actually given by a sum of
the total spin operator and the staggered spin operator
of the model after the transformation.
Thus, the physical absorption spectrum of ESR
is also given by the sum of contributions from the uniform
and staggered part:
\begin{equation}
\chi''_{\rm phys}(q=0,\omega)
    \sim \chi''(q=0,\omega) + \left( \frac{D}{J}\right)^2
 \chi''(q=\pi,\omega) ,
\end{equation}
where $\chi''$ is the imaginary part of the dynamical
susceptibility for the transformed model. The staggered part
$\chi''(q=\pi,\omega)$ is already broad even in the absence of the
anisotropic perturbation, and is further suppressed by the factor
$(D/J)^2$. Thus it would be practically indistinguishable from the
background, especially in the high temperature regime. The main
absorption due to the $\chi''(q=0,\omega)$ term is presumably
Lorentzian with the width given by~(\ref{eq:etaDMelim}). According
to this picture, the lineshape is not a single Lorentzian,
although apparently it is. The direct application of the KT
formula misses such a structure, and treats all the effects as if
the lineshape is a single Lorentzian. This presumably leads to the
incorrect result~(\ref{eq:etaDMdirect}).

An indirect evidence of our claim is that the elimination seems to
work well in the field theory of ESR at low temperature. Assuming
a smooth crossover at $T \sim J$, the latter
result~(\ref{eq:etaDMelim}) seems more plausible. In addition, a
recent experiment\cite{Feyerherm} on a very good one-dimensional
$S=1/2$ Heisenberg antiferromagnet Pyrimidine Cu dinitrate
strongly suggests that there is a staggered DM interaction along
the chain, resulting in the field-induced gap similar to that
observed in Cu benzoate. An analysis of various experimental data
suggests\cite{Feyerherm} that the staggered DM interaction
$D \sim 0.14J$, where the exchange coupling in this compound
is $J \sim 36 {\rm K}$.
According to the direct approach~(\ref{eq:etaDMdirect}), the
linewidth at high temperature should be of the order of
$D^2/J \sim 5000 {\rm Oe}$.
This might be too large to understand the observed small
linewidth $\sim 20 {\rm Oe}$ at room temperature\cite{Feyerherm},
which is quite high compared to the exchange interaction $J$.
On the other hand, if we use eq.~(\ref{eq:etaDMelim}) and
$H \ll D$, the estimate of the linewidth becomes to be of
order of $D^4/J^3 \sim 100 {\rm Oe}$, which is not too far
from the experimental result.
We note that we do not know the numerical coefficients and
thus a conclusive quantitative discussion is difficult.
In addition, the exchange anisotropy (before the elimination of the
DM interaction), which is ignored in the above estimate,
is not known precisely.
Nevertheless, considering the significant difference,
the observed linewidth in pyrimidine Cu dinitrate
could serve as an experimental support for our claim that the direct
treatment of the DM interaction is inappropriate.

On the other hand, we do not understand at present how to deal
with a uniform DM interaction along the chain. While it can be
eliminated by a similar transformation as well, the result
contains the magnetic field rotating in its direction along the
chain. This is a rather unfamiliar problem which we do not know
how to handle at present.

\subsection{High-temperature expansion of the linewidth}

In a series of papers, Yamada and collaborators studied the
temperature dependence of ESR linewidth in one dimensional
magnetic systems experimentally and theoretically. In the
theoretical study, they discussed the temperature dependence by a
high-temperature expansion of the KT formula. More precisely, they
attempted a high-temperature expansion of $\langle \calA
\calA^{\dagger} \rangle$ in the numerator of the KT formula, Eq.
(\ref{eq:KTwidth}).

They concluded that for an exchange anisotropy the linewidth
increases as the temperature is lowered, while the tendency
is the opposite for a (uniform or staggered) DM interaction.
Based on this observation, they argued that the DM interaction
should be dominant in several one-dimensional antiferromagnets
which showed a decreasing linewidth at lower temperature.
In some cases the DM interaction is forbidden according to
the previously identified crystal symmetry; they went on
to the conclusion that the actual symmetry is lower than
what had been believed, allowing the DM interaction.

However, their argument is to be criticized on several grounds.
First, the high-temperature expansion can not be trusted except
for very high temperature. At $T \ll J$ our field theory approach
should be more reliable, and it gives a rather opposite result to
their claim. Second, they expand only the numerator $\langle \calA
\calA^{\dagger} \rangle$ in the KT formula to the first order in
$1/T$, ignoring other possible contributions of order $1/T$. It is
not clear to us whether their scheme makes sense as a $1/T$
expansion of the linewidth. Third, perhaps most importantly, even
in their framework of the calculation, the conclusion should be
reversed because they apparently made a crucial sign mistake as we
will show below. Finally, they apply the KT formula directly to
the DM interaction; this is problematic as we have pointed out. In
any case, the sign problem persists whether the direct approach or
the elimination approach is taken in dealing with the DM term.

In the following, let us show that the sign should be reversed
within the framework of
Refs.~\onlinecite{Yamada:KCuF3,Yamada:DM,Yamada:CuGeO3,Yamada:NaVO}.
We consider
$S=1/2$ Heisenberg antiferromagnetic chains with a small
perturbation. First let us discuss the case of an anisotropy
parallel to the applied field. The calculation for general
anisotropy angle should be similar. This gives the commutator
\begin{equation}
\calA = [ {\cal H}', S^+ ] =
    \delta \sum_j ( S^+_j S^z_{j+1} + S^z_j S^+_{j+1} ) .
\end{equation}
The ``numerator'' of the KT formula is then given by
\begin{equation}
\langle \calA \calA^{\dagger} \rangle
 = \delta^2 \sum_{j,k}
\langle ( S^+_j S^z_{j+1} + S^z_j S^+_{j+1} )
( S^-_k S^z_{k+1} + S^z_k S^-_{k+1} ) \rangle .
\end{equation}
Considering the high-temperature limit
we can ignore all but nearest-neighbor correlations.
Thus we only consider the $j=k$ terms in the double sum;
\begin{equation}
\langle [ {\cal H}', S^+ ] [S^- , {\cal H}' ] \rangle
\sim
\delta^2 N [ \frac{1}{4} + \frac{\langle S^z \rangle}{2}
        + \langle S^x_j S^x_{j+1} \rangle ],
\end{equation}
where we have used identities for $S=1/2$, such as $S^+_j S^-_j =
S^z_j + 1/2$ and $N$ is the number of sites. In the limit of
infinite temperature, the width is given by the first term which
is $\delta^2/J$ as was already discussed.

As the temperature is lowered from infinity, the leading
correction is given by the second and third terms. The second term
proportional to the magnetization $\langle S^z \rangle$ is
negligible compared to the third term in our case $H \ll J$. The
third term represents the nearest-neighbor correlation effect, and
should be proportional to $ - J/T$ at high temperature $T$. Note
that we are dealing with {\em an antiferromagnet}, so that the
nearest-neighbor correlation should be negative. Ignoring other
possible sources of temperature dependence following
Refs.~\onlinecite{Yamada:KCuF3,Yamada:DM,Yamada:CuGeO3,Yamada:NaVO},
the linewidth in the present case is
given by
\begin{equation}
\eta = \frac{{\delta}^2}{J} [ a - b \frac{J}{T}  + O(\frac{J^2}{T^2}) ],
\label{eq:delta-T}
\end{equation}
with positive coefficients $a,b$ for an antiferromagnet. Namely,
the linewidth decreases at lower temperature contrary to the
claims made in
Refs.~\onlinecite{Yamada:KCuF3,Yamada:DM,Yamada:CuGeO3,Yamada:NaVO};
this is rather
natural from the field theory results at low temperatures as
discussed in Section~\ref{sec:hight-anis}.
It appears that, they took
the nearest-neighbor correlation as positive,
which is valid for a ferromagnet\cite{Soos} but not for an antiferromagnet.

Now let us consider the transverse staggered field perturbation
in the same framework.
In this case, $\calA = [ \calH' , S^+ ] = - h \sum_j (-1)^j S^z_j$,
which gives the ``numerator''
\begin{equation}
\langle \calA \calA^{\dagger} \rangle
= h^2 \sum_{j,k} (-1)^{j+k} \langle S^z_j S^z_k \rangle .
\end{equation}
In the high-temperature limit, we may ignore all the correlation
functions other than the nearest neighbor one.
This leads to the  formula
\begin{equation}
\langle \calA \calA^{\dagger} \rangle
= h^2 N [ \frac{1}{4} - 2 \langle S^z_j S^z_{j+1} \rangle ]
\end{equation}
Considering that the nearest-neighbor correlation is negative
for an antiferromagnet, the linewidth is supposed to be given as
\begin{equation}
\eta = \frac{h^2}{J} [ a' +b' \frac{J}{T}  + O(\frac{J^2}{T^2}) ],
\end{equation}
where $a'$ and $b'$ are positive constants.
Namely, the linewidth increases at lower temperature;
again in a qualitative agreement with the field theory.

For a staggered DM interaction, as we have discussed before,
presumably we should first eliminate the DM interaction to
reduce the problem to the exchange anisotropy and the
transverse staggered field.
In a typical situation $H \ll D \ll J$, the staggered field
$h \sim DH/J$ and the anisotropy $\delta \sim D^2/J$
induced by the transformation satisfy $h \ll \delta \ll J$.
In this case, the linewidth would initially decrease by lowering
the temperature, then start increasing below the crossover temperature
where the staggered field becomes dominant.
This was actually observed in Cu benzoate, as discussed already
in Section~\ref{sec:stagexp}.

Finally, we consider a direct application of the KT formula
to the DM interaction. Although we believe this is not an
adequate approach, the claims in
Refs.~\onlinecite{Yamada:KCuF3,Yamada:DM,Yamada:CuGeO3,Yamada:NaVO}
still suffers from the same sign problem even if we accept the direct approach.
Now we have
\begin{equation}
\calA = [{\cal H}' , S^+ ] =
    \sum_j D_j  i (S^z_j S^+_{j+1} - S^+_j S^z_{j+1} ),
\end{equation}
giving
\begin{equation}
\langle \calA \calA^{\dagger} \rangle
    \sim D^2 N [ 1 -  \langle S^x_j S^x_{j+1} \rangle ],
\end{equation}
ignoring other than next-nearest-neighbor correlation in the
high-temperature limit.
Because the nearest neighbor correlation function is negative
in an antiferromagnet, we obtain
\begin{equation}
\eta = \frac{D^2}{J} [ a'' +b'' \frac{J}{T}  + O(\frac{J^2}{T^2}) ],
\end{equation}
where $a''$ and $b''$ are positive constants, implying
the increasing linewidth at lower temperatures.
The error in
Refs.~\onlinecite{Yamada:KCuF3,Yamada:DM,Yamada:CuGeO3,Yamada:NaVO}
is again apparently due
to the identification of the nearest-neighbor correlation as positive.

\section{Conclusions}
\label{sec:conc}

In this paper, we have developed a new approach based on
field theory to ESR in quantum spin chains.
It is expected to be exact in the low-energy (low-temperature) limit,
precisely where the traditional calculational methods on ESR become
invalid.
The weakly broken $SU(2)$ symmetry under an applied field,
in the absence of an anisotropic perturbation,
is represented by the $SU(2)$ symmetric field theory and an anisotropic
mapping between the physical spin operators and the corresponding
field theory operators.

The formulation of the ESR in terms of Feynman-Dyson self-energy
gives, at least in some simple cases, a microscopic derivation of
the Lorentzian lineshape up to a possible smooth weak background. 
The spin diffusion picture\cite{ESRspindiff}, which
predicts a non-Lorentzian lineshape in one dimension, does not
apply to the $S=1/2$ antiferromagnetic chain at low temperature.
The spin diffusion hypothesis does not hold in the present
case, as the spin correlation function is given
explicitly using eq.~(\ref{eq:Schulz}).

The width and shift are calculated
perturbatively for a transverse staggered field perturbation
and an exchange anisotropy parallel or perpendicular to the
applied uniform field.
They seem to explain many existing experimental data.
Furthermore, the self-energy formulation can be used beyond the
perturbation theory. In fact, in the presence of  a 
staggered field, the perturbation theory breaks down at
a low enough temperature.
The ESR spectrum in the zero temperature limit is discussed
with a non-perturbative treatment of the sine-Gordon field theory.
This again seem to explain the experimentally observed
ESR lineshape in Cu benzoate at very low temperature.

While our field theory approach works only at low temperatures,
we have also discussed a few aspects of ESR at higher temperatures.
In particular, we have pointed out that a naive application
of the standard Kubo-Tomita theory fails even in the high temperature
limit, in the presence of a Dzyaloshinskii-Moriya interaction.

We hope that the reader is convinced that
ESR in a strongly interacting quantum system
is quite an interesting problem from the theoretical point of view.
It is also a useful experimental probe because a very precise
spectrum can be obtained.

Obviously, there remain many problems to be investigated in the future.
Even in the simple quantum antiferromagnetic chain,
the formulation of ESR in terms of  self-energy of the boson
field $\phi$ does not hold for generic types of
anisotropic perturbations, because of the mixing of several
operators.
Extension of the self-energy formulation to the generic cases
is an important open problem;
presumably we have to consider perturbative expansion of
correlation functions of the vertex operators (exponentials of the
boson field) in a systematic way.
Moreover, degrees of freedom other than spins
(e.g. charge fluctuation, lattice vibration etc.) will be
relevant in some real materials.
While the ESR in a three-dimensional magnet
appears to be understood with the existing
theory\cite{MK:Ferro,MK:AF}, we think that the problem
should be reinvestigated with the modern understanding
of many-body physics and critical phenomena.
Naturally, the two-dimensional problem, which is expected to
be more sensitive to the fluctuation effects,
would also deserve consideration.
We hope the present work will stimulate further theoretical and
experimental studies on this fascinating subject.

\medskip

{\it Note Added.}
After submitting the present paper,
a paper by Choukroun, Richard and
Stepanov was published\cite{Choukroun}.
They made a similar proposal to ours (Sec.~\ref{sec:hight-DM} in the
present paper) on the treatment of
the Dzyaloshinskii--Moriya interaction at high temperature limit.

\acknowledgments

It is a pleasure to thank Y. Ajiro, J.-P. Boucher, F.~H.~L. Essler,
R. Feyerherm, Y. Maeda, S. Miyashita, Y. Natsume,
H. Nojiri and H. Shiba
for stimulating discussions and useful correspondences.
In particular, we are grateful to D.~M. Edwards for his suggestion
on the content of Appendix and for allowing us to present
it in this paper.
We also thank A.~A. Zvyagin for pointing out Ref.~\cite{Zvyagin}.
This work has been supported in part by a Grant-in-Aid from
MEXT of Japan and from NSERC of Canada.

\appendix

\section{Alternative derivation of the Mori-Kawasaki formula}
\label{sec:deriveMK}

In this appendix, we describe a simple alternative derivation of
the MK formula~(\ref{eq:MKwidth}),(\ref{eq:MKshift}) suggested to
us by David M. Edwards. It depends only on the assumption
that the lineshape takes a single Lorentzian form, and appears
much simpler than that in the original paper\cite{MK:Ferro}.
On the other hand, it does not answer the question why (and when) the
lineshape takes the Lorentzian form.

We consider ESR in a general spin system given by the
Hamiltonian~(\ref{eq:totalH}).
Here and in the following, a spin operator without
a site index is regarded as the total spin operator
$S^{\alpha} = \sum_j S^{\alpha}_j$.
The equations of motion for $S^{\pm}$ are given by
\begin{eqnarray}
    \frac{d S^+}{dt} &=& - i H S^+ + i \calA \\
    \frac{d S^-}{dt} &=& + i H S^- -  i \calA^{\dagger},
\end{eqnarray}
where $\calA = [ \calH' , S^+]$.

The ESR spectrum can be obtained from the Green's function
of $S^{\pm}$.
Let us relate this to the Green's function of
$\calA$ and $\calA^{\dagger}$, using the equations of motion.
Using a partial integration and the equations of motion,
\begin{eqnarray}
\calG^R_{S^+ S^-}(\omega) &=&
-i \int_0^{\infty} e^{i \omega t} \langle [ S^+(t) , S^-(0)] \rangle dt
\nonumber \\
&=&
\frac{1}{\omega} \langle [ S^+(0) , S^-(0)] \rangle
+\frac{1}{\omega} \int_0^{\infty} e^{i\omega t}
    \langle [ \frac{d S^+}{dt}(t), S^-(0)] \rangle dt
\nonumber \\
&=&
\frac{2 \langle S^z \rangle}{\omega}
+ \frac{H}{\omega} \calG^R_{S^+ S^-}(\omega)
- \frac{1}{\omega} \calG^R_{\calA S^-}(\omega) .
\end{eqnarray}
Thus
\begin{equation}
\calG^R_{S^+ S^-}(\omega) =
    \frac{2 \langle S^z \rangle - \calG^R_{\calA S^-}}{\omega -H} .
\label{eq:Gss}
\end{equation}
(Precisely speaking we should introduce the convergence factor
so that $\omega - H$ is replaced by $\omega - H + i \epsilon$
with a positive infinitesimal $\epsilon$.
Although we omit this for brevity, it can be recovered
when necessary.)
Performing similar steps,
\begin{eqnarray}
\calG^R_{\calA S^-}(\omega) &=&
-i \int_0^{\infty} e^{i \omega t} \langle [ \calA(t) , S^-(0)] \rangle dt
\nonumber \\
&=&
\frac{1}{\omega} \langle [ \calA(0) , S^-(0)] \rangle
- \frac{1}{\omega} \int_0^{\infty} e^{i \omega t}
    \langle [ \calA(t) , \frac{d S^-}{dt}(0)] \rangle dt
\nonumber \\
&=&
+ \frac{\langle [\calA(0), S^-(0) ] \rangle}{\omega}
+ \frac{H}{\omega} \calG^R_{\calA S^-}(\omega)
- \frac{1}{\omega} \calG^R_{\calA \calA^{\dagger}}(\omega),
\end{eqnarray}
where we used the relation
\begin{equation}
 \frac{d}{dt} \langle [ \calA(t), S^-(0)]\rangle =
- \langle [\calA(t), \frac{d S^-}{dt}(0) ] \rangle
\end{equation}
which holds because the Green's function depends only
on the difference of two time arguments. Thus
\begin{equation}
\calG^R_{\calA S^-}(\omega) =
\frac{
\langle [ \calA(0), S^-(0) ] \rangle
- \calG^R_{\calA \calA^{\dagger}}(\omega)}{\omega -H} .
\label{eq:Gas}
\end{equation}
Combining eqs.~(\ref{eq:Gss}) and (\ref{eq:Gas}), we obtain
\begin{equation}
\calG^R_{S^+ S^-}(\omega) =
    \frac{ 2 \langle S^z \rangle}{\omega -H}
+ \frac{ - \langle [ \calA(0), S^-(0) ] \rangle
+ \calG^R_{\calA \calA^{\dagger}}(\omega)}{(\omega -H)^2} .
\label{eq:Gss2}
\end{equation}
This should be an exact relation between the full Green's functions
(in which the effect of the perturbation $\calH'$ is fully
taken into account.)
When $\calH'=0$, we recover the simple result
$\calG^R_{S^+ S^-}(\omega) = 2 \langle S^z \rangle / (\omega -H)$.

Now let us assume the perturbation $\calH'$ is small, and the ESR
lineshape is Lorentzian. Namely, we assume that
$\calG^R_{S^+S^-}(\omega)$ is given by eq.~(\ref{eq:lor}) where $\Sigma$ is a
smooth function of $\omega$. Near the resonance $\omega \sim H$,
$\Sigma$ may be regarded as a constant. ${\rm Re} \Sigma$ and $-
{\rm Im} \Sigma$ gives the shift and width of the resonance,
respectively. We assume that $\Sigma$ can be expanded
perturbatively in $\calH'$.

Comparing eqs.~(\ref{eq:lor}) and (\ref{eq:Gss2}), we obtain,
{\em in the lowest order of perturbation theory},
\begin{equation}
\Sigma \sim
\frac{ - \langle [ \calA(0), S^-(0) ] \rangle
    + \calG^R_{\calA \calA^{\dagger}}(\omega=H)}{2 \langle S^z \rangle}.
\end{equation}
Here we note that $\langle [ \calA(0), S^-(0) ] \rangle$ is purely
real since $[\calA,S^-]$ is Hermitean.
This gives
\begin{eqnarray}
\eta &=& \frac{-1}{2 \langle S^z \rangle}
    {\rm Im}\calG^R_{\calA \calA^{\dagger}}(\omega=H), \\
\Delta \omega &=& \frac{1}{2 \langle S^z \rangle}
    \left[  -\langle [ \calA, S^- ] \rangle
     + {\rm Re}\calG^R_{\calA \calA^{\dagger}}(\omega=H) \right] .
\end{eqnarray}
For a small field $H$,
the denominator $2 \langle S^z \rangle$ can be written
as $2 \chi_u H$ where $\chi_u$ is the uniform susceptibility.
We also note that the first term in the shift
$-\langle [ \calA, S^-] \rangle/(2\langle S^z \rangle)$
was derived previously by Kanamori and Tachiki\cite{KanamoriTachiki}, and by
Nagata and Tazuke\cite{NT}.
However, their theory did not incorporate the dynamical
effects represented by $\calG^R_{\calA \calAdag}$.

So far, we have defined the expectation value and
the Green's functions with respect to the
full Hamiltonian $\calH = \calH_0 + \calH_Z + \calH'$.
However, the present result is only valid in the leading order.
Since the Green's function of $\calA$ above already contains
the factor $\lambda^2$ ($\lambda$ is the small parameter that
characterizes the perturbation $\calH'$), we may
replace $\calG^R_{\calA \calAdag}$
with the unperturbed Green's function $G^R_{\calA \calAdag}$.
This gives the formulae in eqs.~(\ref{eq:MKwidth}) and (\ref{eq:MKshift}).
We note that, in general,
$\langle [\calA,S^-]\rangle$ must still be evaluated
in the presence of $\calH'$ because $[\calA,S^-]$ is only
first order in $\lambda$.

\end{document}